\documentclass{article}
\usepackage[english]{babel}
\usepackage{cite}
\usepackage{bbold}
\usepackage[makeroom]{cancel}
\usepackage{soul}
\usepackage{subcaption}
\usepackage{url}
\usepackage{authblk}

\usepackage[letterpaper,top=2cm,bottom=2cm,left=3cm,right=3cm,marginparwidth=1.75cm]{geometry}

\usepackage{amsmath}
\usepackage{graphicx}
\usepackage[colorlinks=true, allcolors=blue]{hyperref}   

\title{Moyal deformation of the classical arrival time}

\author[1,2]{Dean Alvin L. Pablico \footnote{dlpablico@up.edu.ph}}
\author[2]{Eric A. Galapon\footnote{eagalapon@up.edu.ph}}
\affil[1]{\footnotesize University of Northern Philippines, 2700 Vigan City, Ilocos Sur, Philippines}
\affil[2]{\footnotesize Theoretical Physics Group, National Institute of Physics\\University of the Philippines, Diliman, 1101 Philippines}

\begin{document}
\maketitle

\begin{abstract}
The quantum time of arrival (TOA) problem requires the statistics of measured arrival times given only the initial state of a particle. Following the standard framework of quantum theory, the problem translates into finding an appropriate quantum image of the classical arrival time $\mathcal{T}_C(q,p)$, usually in operator form $\hat{\mathrm{T}}$. In this paper, we consider the problem anew within the phase space formulation of quantum mechanics. The resulting quantum image is a real-valued and time-reversal symmetric function $\mathcal{T}_M(q,p)$ in formal series of $\hbar^2$ with the classical arrival time as the leading term. It is obtained directly from the Moyal bracket relation with the system Hamiltonian and is hence interpreted as a Moyal deformation of the classical TOA. We investigate its properties and discuss how it bypasses the known obstructions to quantization by showing the isomorphism between $\mathcal{T}_M(q,p)$ and the rigged Hilbert space TOA operator constructed in  [Eur. Phys. J. Plus \textbf{138}, 153 (2023)] which always satisfy the time-energy canonical commutation relation (TECCR) for arbitrary analytic potentials. We then examine TOA problems for a free particle and a quartic oscillator potential as examples.
\end{abstract}

\section{Introduction}
The incorporation of time as a quantum dynamical observable, dubbed the quantum time problem (QTP), has remained unsolved since the birth of quantum mechanics. Diverse theoretical approaches have been proposed, but a consensus has yet to be reached \cite{Muga2008, Muga2009}. In the standard formulation of quantum theory, time is only considered an external parameter, just as in classical mechanics. However, the interpretation of the time-energy uncertainty principle \cite{Galapon2002} and questions involving the occurrence and duration of quantum processes, e.g. decay, dwell, transitions, arrivals, and tunneling, require a dynamical treatment of time \cite{Galapon2001}. The advent of experimental attosecond ionization techniques makes the problem more important than ever as theories involving time can now be tested up to a quintillionth of a second ($10^{-18}$s) \cite{Eckle2008,Sainadh2019,Eckle2008a}.

To gain a better perspective on the quantum time problem, the physics community often considers the arrival time of an elementary particle due to its conceptual simplicity \cite{Allcock1969a,Allcock1969b,Allcock1969,Grot1996,Aharonov1998,Leavens1998,Delgado1998,Wang2007,Muga2000,J.Leon20000,Galapon2004,Galapon2018,Galapon2009a,Galapon2006,Halliwell2015,Pollak2017,Sombillo2018,Galapon2004a,Anastopoulos2006,Sombillo2016,Das2021,Galapon2002a,Galapon2005,Galapon2005a,Galapon2008,Caballar2009,Caballar2010,Villanueva2010,Flores2019,Galapon2009,Galapon2012,Pablico2020, Sombillo2014, Baute2000}. The problem proceeds as follows \cite{J.Leon20000,Galapon2012}. Let us consider an ideal experiment with a structureless particle of mass $\mu$, initial momentum $p$ and initial position $q$, with the aim to measure its arrival time $t=T_x(q,p)$ at a specific point $q(t=T_x)=x$, say at the origin $x=0$. We place a detector $D_T$ at the origin to signal the particle's arrival, and another detector $D_R$ to its far left. An interaction potential $V(q)$ is prepared in between these two detectors to study any position-dependent interaction effects between the particle and its medium. Initially, the incident particle is prepared in a state $\psi(q)$ of the form, $\psi(q) = \varphi(q) e^{ik_0 q}$, where $\varphi(q)$ satisfies  $\int_{-\infty}^\infty dq \,\bar{\varphi}(q) \,\varphi'(q)$  =0, $e^{ik_0 q}$ is a momentum eigenstate, and $k_0=\sqrt{2\mu E_0}/\hbar$ for a given incident energy $E_0.$ The imposed condition on $\varphi(q)$ assures that $\psi(q)$ has only a momentum dependence of the factor $e^{ik_0 q}$. The particle is then launched at time $t=0$. The arrival time of the particle at the origin is recorded when $D_T$ clicks, while no arrival time is measured when $D_R$ clicks. Now, if we repeat the same experiment several times with the same initial state for every repeat, how do we obtain the statistics of measured arrival times at the arrival point? 

The proposed measurement scheme provides the ideal statistics for measured arrival times. Despite this, it remains relevant because these ideal statistics can, in principle, be mapped into another positive operator-valued measure (POVM) depending on specific experimental details. This means that the obtained statistics from a particular measurement model can be derived through smearing or post-processing of the ideal statistics obtained from the above measurement model. This smearing process entails the construction of a POVM, denoted as $\textbf{P}$, which describes an imperfect measurement of the time of arrival observable $\mathcal{T}$. However, constructing such a POVM requires a separate and more detailed study, which we do not address in this paper. 

In classical mechanics, the problem is easily solved using the classical arrival time expression given by
\begin{equation}\label{classical}
\mathcal{T}_C(q,p)=-\mathrm{sgn}(p)\sqrt{\frac{\mu}{2}}\int_{0}^{q} \, \frac{dq'}{\sqrt{H(q,p)-V(q')}},
\end{equation}
where $\mathrm{sgn}(p)$ is the signum function. Equation (\ref{classical}) is obtained simply by inverting the particle's equation of motion. Note, however, that it is not defined in the entire phase space. It is real and finite only in classically allowed regions. For potentials like a harmonic oscillator, it can become multiple-valued, indicating multiple arrivals at the arrival point. In cases of quantum tunneling across a potential barrier, it can become complex-valued, implying non-arrival of the classical particle at the arrival point. In instances where the classical arrival time is negative, like in the case of a free particle with $q<0$ and $p<0$, the particle moves leftward, away from the designated arrival point at the origin. In our measurement scheme, the detector $D_R$ will register a click and hence, no arrival time is measured.

Now, if we follow the usual prescription for observables in quantum theory, the quantum TOA problem translates into finding an appropriate \textit{quantum image} of the classical arrival time (\ref{classical}). This quantum image serves as a mathematical tool for extracting expectation values and probability distributions subject to experimental comparisons. But this begs a crucial question: How do we construct such a quantum image?

In the standard Hilbert space formulation of quantum mechanics, quantum images of classical observables are represented by hermitian operators in the underlying state space. But the continuous spectrum of time observables and the possible singularity of their eigenfunctions require the extension of quantum theory to the rigged Hilbert space (RHS) \cite{Madrid2005, Madrid2002, Galapon2004,Galapon2018,Pablico2023}. In this framework, TOA operators generally appear as integral operators of the form
\begin{equation}\label{operatordef}
(\hat{\mathrm{T}}\,\varphi )(q)=\int_{-\infty}^{\infty} \,dq' \,\langle q|\hat{\mathrm{T}}|q'\rangle \,\varphi (q'),
\end{equation}
in configuration space. In Ref. \cite{Galapon2018}, quantized time of arrival (QTOA) operators $\hat{\mathrm{T}}_Q$ of the form given by Eq. (\ref{operatordef}) are constructed using Weyl, Born-Jordan, and symmetric quantizations of the classical arrival time. The kernels of the operators all appear as
\begin{equation}\label{tkfdef}
\langle q|\hat{\mathrm{T}}_Q|q'\rangle=\frac{\mu}{ih}T_Q(q,q')\,\mathrm{sgn}(q-q'),
\end{equation}
where the explicit form of the kernel factor $T_Q(q,q')$ depends on which quantization rule $Q$ is used. Quantum images constructed this way, at least for the three quantizations discussed above, are hermitian, $\hat{\mathrm{T}}_Q=\hat{\mathrm{T}}_Q^\dagger$, and time-reversal symmetric, $\Theta \hat{\mathrm{T}}_Q \Theta ^{-1}=-\hat{\mathrm{T}}_Q$, where $\Theta$ is the time reversal operator. The latter property has a direct role on the dynamics of time of arrival operators. For the free particle case, quantized operators with time-reversal symmetry have eigenfunctions with ideal unitary arrival properties, where their probability density at the arrival point, either maximizes (indicating particle detection) or vanishes (indicating non-detection) at a time equal to their corresponding time eigenvalue. Without time-reversal symmetry, operators lose this dynamical property \cite{Caballar2010}. 

They also satisfy correctly the quantum-classical correspondence in the classical limit $\hbar \to 0$. Nevertheless, quantized TOA operators generally fail to satisfy the required conjugacy with the system Hamiltonian, called the time-energy canonical commutation relation (TECCR), given by
\begin{equation}\label{teccr}
[\hat{\mathrm{H}},\hat{\mathrm{T}}]=i\hbar \mathbb{1},
\end{equation}
for arbitrary analytic potentials. Only the Weyl QTOA operator satisfies Eq. (\ref{teccr}) but is limited to linear systems that are up to quadratic in position \cite{Galapon2001}. The general insufficiency of the theory of quantized TOA operators is often attributed to the known obstructions to quantization \cite{Groenewold1946,Hove1951,Gotay1996,Gotay1999}, which rejects the notion of quantizing every classical polynomial observable consistently in phase space so that the Poisson-bracket-commutator correspondence is always satisfied. It is then clear that canonical quantization is inadequate if one requires the correct algebra of time observables in accordance to Eq. (\ref{teccr}). 

An alternative approach has been introduced in Ref. \cite{Galapon2004} independent of canonical quantization, called supraquantization. The basic idea is to construct quantum images of the classical arrival time directly from the axioms of quantum mechanics and the known properties and algebra of time observables. Here, the classical observable only serves as a boundary condition, not the starting point of calculations, unlike canonical quantization. The constructed quantum image, the supraquantized TOA operator $\hat{\mathrm{T}}_S$, and its kernel $\langle q|\hat{\mathrm{T}}_S|q'\rangle$ also share the same general form as Eqs. (\ref{operatordef}) and (\ref{tkfdef}), respectively. The difference, however, lies on the kernel factor $T_S(q,q')$ which appears as a solution of the following second-order partial differential equation,
\begin{equation}\label{TKE2}
-\frac{\hbar^2}{2\mu} \frac{\partial^2 T_S(q,q')}{\partial q^2}+\frac{\hbar^2}{2\mu} \frac{\partial^2 T_S(q,q')}{\partial q'^2}+ \left[V(q)-V(q')\right]T_S(q,q')=0,
\end{equation}
called the time kernel equation (TKE) subject to the boundary conditions, $T_S(q,q)=q/2$ and $T_S(q,-q)=0.$ The time kernel equation arises by imposing the canonical commutation relation (\ref{teccr}) on the operator $\hat{\mathrm{T}}_S$. It admits a unique solution for entire analytic interaction potentials \cite{Sombillo2012,Farrales2022}.

As shown recently in Ref. \cite{Pablico2023}, the solution of the TKE admits the expansion
\begin{equation}
T_S(q,q')=\sum_{n=0}^{\infty}T_{S,n}(q,q'),
\end{equation} 
where $T_{S,0}(q,q')$ coincides with the Weyl map of the classical arrival time's expansion around the free-particle arrival time. The succeeding terms $T_{S,n}(q,q')$ are obtained recursively from
\begin{equation}\label{tnuvintro}
\begin{split}
T_{S,n}(q,q')=&\left(\frac{\mu}{2\hbar^2}\right)\sum_{r=1}^{n} \frac{1}{(2r+1)!}\frac{1}{2^{2r}}\int_{0}^{q+q'} ds \, V^{(2r+1)}\left(\frac{s}{2}\right) \int_{0}^{q-q'} dw \, w^{2r+1} \, T_{S,n-r}(s,w)\,G(s,w),\\
\end{split}
\end{equation}
where
\begin{equation}\label{gsw}
G(s,w)={}_0F_1 \left(;1;\left(\frac{\mu}{2\hbar^2}\right)((q-q')^2-w^2)\left[V \left(\frac{q+q'}{2}\right)-V \left(\frac{s}{2}\right)\right]\right),
\end{equation}
with $T_{S,0}(q,q')$ being the initial condition. The operator $\hat{\mathrm{T}}_S$ then appears as an infinite series of integral operators with the leading term as the Weyl quantization of the classical arrival time $\hat{\mathrm{T}}_W$, i.e., $\hat{\mathrm{T}}_S=\hat{\mathrm{T}}_W+\hat{\mathrm{T}}_1+\hat{\mathrm{T}}_2...$. The succeeding terms, $\hat{\mathrm{T}}_n$ for $n\ge1$, have been interpreted as the quantum corrections to $\hat{\mathrm{T}}_W$ in rigged Hilbert space. These quantum corrections vanish for linear systems so that the supraquantized TOA operator coincides with the Weyl-quantized TOA operator for such systems. 

Now, the theory of supraquantization is, in fact, not the only possible alternative path to quantization. There is also the more well-known phase space formulation of quantum mechanics, or simply the quantum phase space (QPS), jointly formulated by Weyl, Wigner, and Moyal (WWM) \cite{Zachos2001,Zachos2005}. It is a logically complete and self-standing formulation of quantum mechanics and has been proven to be useful to all areas of physics, including quantum optics, nuclear physics, quantum information, collision theory, and non-linear physics, among others \cite{HaiWoongLee1995}. 

The QPS is based on (i) Weyl's correspondence between Hilbert space operators and phase-space functions, (ii) Wigner's quasi-probability distribution, and (iii) Moyal's $\star$-product and bracket between two phase-space functions \cite{Zachos2005}. In this framework, the symplectic phase space of classical mechanics is deformed into a non-commutative phase space generated by the position and momentum observables. This makes it ideal for the quantum TOA problem where one considers the non-commutative algebra of time observables with respect to the system Hamiltonian. It also allows canonical transformations similar to classical mechanics, which may provide a deeper understanding of the uncertainty principle \cite{Kim1991}. In addition, it places the position and momentum observables on equal footing so that the QPS may offer a more universal approach to the quantum TOA problem. Conversion to position or momentum representations can be easily done simply by integration over the other space.  

We are then motivated to explore the application of the QPS formulation to the quantum TOA problem, which has yet to be done in the literature. Specifically, we aim to (i) introduce an alternative framework for obtaining quantum images of the classical arrival time independent of quantization and operator-based formulations, (ii) test the theoretical predictions of the theory of supraquantization by comparison with that of the current phase space approach, and (iii) identify possible modifications to the theory of quantized TOA operators if one wishes to incorporate the correct algebra of time observables.

Relevant to our objectives is the fact that in phase space quantization, the Moyal bracket, not the Poisson bracket, corresponds to the commutator relation between two operators \cite{Zachos2001,Zachos2005,Jordan1961}. This allows us to recast the algebra of time observables in operator formulation (\ref{teccr}) as the Moyal bracket between two $c$-number functions
\begin{equation}\label{moyalll}
\big\{H(q,p),\mathcal{T}_M(q,p)\big\}_{MB}=1,
\end{equation}
where $H(q,p)$ is the system Hamiltonian and $\mathcal{T}_M(q,p)$ is the required TOA phase space function. In essence, this paper generally revolves around the construction and investigation of $\mathcal{T}_M(q,p)$. Specifically, we will show that it is a real-valued and time-reversal symmetric function in formal series in the deformation parameter $\hbar^2$. The leading term coincides with the classical arrival time and the succeeding terms are interpreted as quantum corrections in phase space. We then determine the corresponding operator in rigged Hilbert space and show that $\mathcal{T}_M(q,p)$ leads exactly to the supraquantized TOA operator $\hat{\mathrm{T}}_S$ constructed before independently of canonical quantization and phase space methods \cite{Galapon2004,Pablico2023}. 
This result proves the isomorphism between the two approaches within the quantum TOA problem leading essentially to the same theoretical predictions. In our context, we refer isomorphism to the direct one-to-one mapping or correspondence between time of arrival observables in the quantum phase space formulation and the theory of supraquantization defined in the rigged Hilbert space formulation. This mapping preserves the essential properties of our arrival time observables, ensuring a consistent description of the same physical quantities across both mathematical frameworks. 

This work is organized as follows. Section (\ref{sec:PB}) reviews the Poisson bracket between time and energy in classical phase space. A derivation of the classical arrival time from the required classical algebra is presented. In Section (\ref{sec:mbtoa}), we construct the quantum image of the classical arrival time directly from the Moyal bracket relation with the system Hamiltonian. Its properties are also discussed. Section (\ref{sec:weylmap}) provides a comparison between the quantum phase space approach and the theory of supraquantization by Weyl correspondence. Section (\ref{sec:examples}) explores two examples including the time of arrival for a vanishing potential and for a quartic oscillator potential. Finally we provide brief concluding remarks in Section (\ref{sec:conclusions}).

\section{The Poisson bracket and the classical arrival time}\label{sec:PB}

Let us first consider a classical structureless point particle initially located at some specific point ($q(t=0),p(t=0)$) in phase space. Its dynamics is dictated by the Hamiltonian 
\begin{equation}
H(q,p)=\frac{p^2}{2\mu}+V(q),
\end{equation}
for a given interaction potential $V(q)$. The arrival time corresponding to the first crossing at the arrival point is simply described by Eq. (\ref{classical}) which holds for right or left moving classical particle. 

A more rigorous derivation of the classical expression $\mathcal{T}_C(q,p)$, including the existence conditions and its characterization as a function of the phase space variables $q$ and $p$, have been provided by Leon, et al. in Ref. \cite{J.Leon20000}. They showed that the Hamilton equations define an integrable flow with a system of holonomic coordinates $(q(t),p(t))$ in phase space for each instant of time, i.e., $q(t)=q(q_0,p_0;t)$, and $p(t)=p(q_0,p_0;t)$. This suggests that the system arrives at a point $(q(t),p(t))$ given a set of initial coordinates $(q_0,p_0)$. These points can then be used to define the arrival time and hence qualified as a derived variable in phase space. From the Hamilton-Jacobi equation, one arrives at Eq. (\ref{classical}). 

However, there is another derivation of $\mathcal{T}_C(q,p)$ which is dependent on the conjugacy relation with the system Hamiltonian $H(q,p)$ in \textit{classical} phase space. To start, recall that a classical particle moves along trajectories defined by $H(q,p)=E=\mathrm{constant}$ as time $t$ increases \cite{J.Leon20000}. Therefore, we need to understand how $\mathcal{T}_C(q,p)$ evolves with respect to the parametric time $t$. We derive this condition on physical grounds.

Consider a particle starting at position $q = q_0$ at time $t = 0$. If the time of arrival at the origin $q = 0$ is $\mathcal{T}_C(q_0, p_0) = 5 \,\mathrm{s}$, then $1\, \mathrm{s}$ after the particle leaves $q_0$, it will be at position $q_1$ and the arrival time will be $\mathcal{T}_C(q_1, p_0) = 4\, \mathrm{s}$. After $2\, \mathrm{s}$, the particle which is at position $q_2$, will arrive in $\mathcal{T}_C(q_1,p_0)=3 \,\mathrm{s}$. After $3\, \mathrm{s}$, the particle will reach the arrival point in $2 \,\mathrm{s}$, and so on. This suggests that $\mathcal{T}_C(q,p)$ should decrease in step with the parametric time $t$ according to $d\mathcal{T}_C(q,p)/dt = -1.$ From Liouville’s equation, this evolution suggests the following classical conjugacy relation
\begin{equation}\label{pb}
\big\{H(q,p),\mathcal{T}_C(q,p)\big\}_{PB}=1.
\end{equation}
The derived Poisson bracket determines how $\mathcal{T}_C(q,p) $ changes under a transformation generated by the Hamiltonian $H(q,p)$. 

We can expand Eq. (\ref{pb}) to obtain the following partial differential equation
\begin{equation}\label{pdeclass}
\frac{\partial V(q)}{\partial q}\frac{\partial \mathcal{T}_C(q,p)}{\partial p}-\frac{p}{\mu}\frac{\partial \mathcal{T}_C(q,p)}{\partial q}=1. 
\end{equation}
In the absence of a potential, i.e., $V(q)=0$, the well-known free-particle arrival time at the origin $\mathcal{T}_F(q,p)=-\mu q/p$ readily emerges. We rewrite Eq. (\ref{pdeclass}) as an integral equation of the form
\begin{equation}\label{succtqp}
\mathcal{T}_C(q,p)=-\frac{\mu q}{p}+\frac{\mu}{p}\int_{0}^{q} dq'\frac{\partial V(q')}{\partial q'}\frac{\partial \mathcal{T}_C(q',p)}{\partial p},
\end{equation}
whose solution can be obtained by the method of successive approximation. We can choose the first term as our zeroth-order approximation of Eq. (\ref{succtqp}), i.e., $\mathcal{T}_{C,0}(q,p)= -\mu q/p$ so that the boundary condition, $\mathcal{T}_F(q,p)$, immediately emerges in the limit of vanishing potential. The $n$th-order approximation for $ n\ge 1$ can be obtained from the recurrence relation
\begin{equation}\label{succtqpn}
\mathcal{T}_{C,n}(q,p)=-\frac{\mu q}{p}+\frac{\mu}{p}\int_{0}^{q} dq'\frac{\partial V(q')}{\partial q'}\frac{\partial \mathcal{T}_{C,n-1}(q',p)}{\partial p},
\end{equation}
with $\mathcal{T}_{C,0}(q,p)$ serving as the initial condition. Comparing Eqs. (\ref{succtqp}) and (\ref{succtqpn}), we identify the relation $\mathcal{T}_C(q,p)=\lim_{n\to\infty}\mathcal{T}_{C,n}(q,p)$. 

The solution of Eq. (\ref{succtqpn}) is given in Appendix (\ref{appendix1}) so that in the limit $n\to \infty$, we arrive at
\begin{equation}\label{ltoapb}
\tau_{M,0}(q,p)=\lim_{n\to\infty}\mathcal{T}_{C,n}(q,p)=-\sum_{k=0}^{\infty}(-1)^k \frac{(2k-1)!!}{k!} \frac{\mu^{k+1}}{p^{2k+1}}\int_{0}^{q}dq' \, \left(V(q)-V(q')\right)^k.
\end{equation}
The above infinite series converges absolutely and uniformly to the classical arrival time (\ref{classical}) when $|V(q)-V(q')|<p^2/2\mu.$ Hence, $\mathcal{T}_C(q,p)$ is derived. We highlight the differences between the expansion $\tau_{M,0}(q,p)$ and the classical arrival time $\mathcal{T}_{C}(q,p).$ In the theory of TOA operators, $\tau_{M,0}(q,p)$ is called the local time of arrival (LTOA) while $\mathcal{T}_{C}(q,p)$ as the global time of arrival (GTOA) \cite{Galapon2004,Galapon2018}. The classical time of arrival holds in the entire region $\Omega=\Omega_q \times \Omega_p$, while the local time of arrival  $\tau_{M,0}(q,p)$ holds only in some local neighborhood of $\omega_q$ of $\Omega_q$ where it is real-valued. This defines the inclusion $\tau_{M,0}(q,p) \subset \mathcal{T}_{C}(q,p)$. Hence, $\mathcal{T}_{C}(q,p)$ is the analytic continuation of  $\tau_{M,0}(q,p)$ in $\Omega/\omega$, where $\omega=\omega_q \times \omega_p$. 

For the case $p \neq 0$ and $V(q)$ continuous at $q$, there exists a neighborhood of $q$ determined by $|V(q)-V(q')|< p^2/2\mu$ such that $\tau_{M,0}(q,p)$ converges absolutely and uniformly to  $\mathcal{T}_{C}(q,p)$, i.e. $\mathcal{T}_{C}(q,p) =\tau_{M,0}(q,p)$ in the region $\omega \subset \Omega$. This is exactly the case when we are able to close the above infinite series as the classical arrival time. Outside the region $\omega_q$, the LTOA $\tau_{M,0}(q,p)$ diverges implying non-arrival at the arrival point. An example of a diverging LTOA is the time of arrival involving a tunneling quantum particle where it has no classical counterpart. That is, in the presence of a potential barrier, a classical particle with insufficient energy can never arrive at the arrival point.

In the general theory of quantized TOA operators, canonical quantization is done on the LTOA \cite{Galapon2001,Galapon2008,Galapon2018}. This is done because, unlike the classical TOA where it can be complex and multiple valued, the LTOA is single and real-valued within its entire region of convergence in phase space. The resulting TOA operator is still considered as the quantum image of the classical arrival time since the latter is the analytic continuation of the former. 

\section{The Moyal bracket and the quantum image of the classical arrival time }\label{sec:mbtoa}

By Dirac's correspondence between Poisson brackets in classical mechanics and commutator brackets in quantum mechanics, Eq. (\ref{pb}) corresponds to the canonical commutation relation $[\hat{\mathrm{H}},\hat{\mathrm{T}}]=i\hbar \mathbb{1}$. However, the existence of obstructions to quantization already rejects this correspondence for any arbitrary functions, except for linear functions in phase space \cite{Groenewold1946,Hove1951,Gotay1996,Gotay1999}. 

In their seminal works, Moyal and Groenewold have shown how to formulate quantum mechanics in a more natural and logical manner using real-valued functions on phase space \cite{Moyal1949a,Groenewold1946}. In doing so, they found that quantum commutators correspond to a more complicated type of bracket called the Moyal bracket, and not the Poisson bracket. The Moyal bracket has the properties of a Lie bracket and is interpreted as a generalization and deformation of the Poisson bracket. Phase-space functions then form a Lie algebra with this bracket \cite{Jordan1961} so that the Moyal bracket algebra is the unique one-parameter deformation of the Poisson bracket algebra \cite{Zachos2005}. The operator representation of this algebra provides the usual formulation of quantum mechanics. 

Moyal and Groenewold's success has led to the development of deformation quantization implying quantum mechanics as a theory of functions and distributions in phase space endowed with deformed products and brackets \cite{Bayen1978}. This means that one can view quantum mechanics as a formal deformation of classical mechanics \cite{Bayen1977}.

In our quest to obtain an algebra-preserving quantum image of the classical arrival time, it then seems natural to use the Moyal bracket instead of quantizing the time observable from the Poisson bracket.  We do this in the current section. We begin our analysis by imposing the known properties of quantum observables and specific features of arrival time functions.

\subsection{General form of the algebra-preserving TOA phase space function}

Our main goal is to solve the phase space TOA function $\mathcal{T}_M(q,p)$ that satisfies the canonical conjugacy relation $\big\{H(q,p),\mathcal{T}_M(q,p)\big\}_{MB}=1$. For convenince, we call $\mathcal{T}_M(q,p)$ as the Moyal time of arrival. We start by assuming the following general form in formal series in $\hbar$,
\begin{equation}\label{phasespacet}
\mathcal{T}_M(q,p)=\sum_{n=0}^{\infty}\hbar^n\,\mathcal{T}_{M,n}(q,p),
\end{equation}
where each $\mathcal{T}_{M,n}(q,p)$ is real, independent of $\hbar$ and satisfies the condition 
\begin{equation}\label{bcnegn}
  \mathcal{T}_{M,n}(q,p)=0  ; \,\, \,\,\,\,\text{for all $n<0$.} 
\end{equation}
The leading term is simply $\tau_{M,0}(q,p) \subset\mathcal{T}_C(q,p)$ so that the Moyal arrival time $\mathcal{T}_M(q,p)$ immediately reduces to the classical arrival time in the limit of vanishing $\hbar$, thus preserving the correspondence between quantum and classical observables. 

Similar form as Eq. (\ref{phasespacet}) has already been postulated before by Tosiek and Brzykcy as the general form of quantum observables represented by real, albeit smooth, phase space functions \cite{Tosiek2012}. A similar expansion has been used by Blaszak and Domanski in their construction of \textit{quantum} Hamiltonians in Ref. \cite{MaciejBlaszak2012}. We are extending their results to time of arrivals wherein we may encounter singularities for some specific values of $q$ or $p$. The divergences, if encountered, may be safely manipulated in the distributional sense.

Next important property we impose on $\mathcal{T}_M(q,p)$ is time-reversal symmetry. We argue that theoretical predictions obtained from $\mathcal{T}_M(q,p)$ should still be consistent and physically significant even if events run backwards in time or all motions are reversed. For example, we consider the simplest case where a particle travels in a free region from the $-q$ axis to the origin. The corresponding arrival time is described by $\mathcal{T}_F(q,p)=-\mu q/p$ so that when $q=q_0$ and $p=p_0$, we simply arrive at the classical time of flight of a free particle. Now, the same measurement should be extracted in the case where both the motion of the particle and the starting and arrival points are reversed. This happens when $\mathcal{T}_F(q,p)$ satisfies the condition $\mathcal{T}_F(q,p)=-\mathcal{T}_F(q,-p)$, which is indeed the case. 

In fact, time-reversal symmetry appears for the general case where an interaction potential $V(q)$ is present. This is already apparent from the classical and local arrival times, $\mathcal{T}_C(q,p)$ and $\tau_{M,0}(q,p)$, given by Eqs. (\ref{classical}) and (\ref{ltoapb}), respectively. Since both of these TOA observables are of the same class as the general observable $\mathcal{T}_M(q,p)$, with the former from the limit $V(q)\to 0$ and the latter from $\hbar \to 0$, $\mathcal{T}_M(q,p)$ should satisfy the same symmetry condition
\begin{equation}\label{trversalsym}
\mathcal{T}_M(q,p)=-\mathcal{T}_M(q,-p).
\end{equation}

In essence, we have exploited the so-called transfer principle introduced in Ref. \cite{Galapon2004}. It suggests that each element of a class of observables shares a common set of properties with the rest of its class. Thus, we can \textit{transfer} a particular property of a specific element of the class to the rest without discrimination. Since we have identified that time-reversal symmetry is satisfied by the free and classical interacting cases, we are assured that the same symmetry condition holds true for other TOA observables of the same class, such as $\mathcal{T}_M(q,p)$. It is an open problem whether or not time-reversal symmetry is satisfied by all class of time observables.

An immediate consequence of time-reversal symmetry is that $\mathcal{T}_M(q,p)$ and its expansion around the free TOA are odd in $p$. This is already evident in the three TOAs, $\mathcal{T}_F(q,p)$, $\mathcal{T}_C(q,p)$, and $\tau_{M,0}(q,p)$. Among all possible phase space functions that may satisfy the Moyal bracket relation with the system Hamiltonian, time-reversal symmetry chooses which among them is allowed as a TOA phase space function. 

\subsection{Explicit calculation of the phase space TOA function}

We now proceed to solve $\mathcal{T}_M(q,p)$. The full expansion of the required Moyal bracket is given by
\begin{equation}\label{moyalht}
1=\big\{H(q,p),\mathcal{T}_M(q,p)\big\}_{MB}=\sum_{r=0}^{\infty}\,\left(\frac{\hbar}{2}\right)^{2r}\frac{(-1)^r}{(2r+1)!} \,\left[ \frac{\partial}{\partial q_H}\, \frac{\partial}{\partial p_\mathcal{T}}-\frac{\partial}{\partial p_H}\,\frac{\partial}{\partial q_\mathcal{T}}\right]^{2r+1}H(q,p)\,\mathcal{T}_M(q,p),
\end{equation}
where the differential operators $\partial/\partial q_H$ and $\partial/\partial p_H$ act only on the Hamiltonian $H(q,p)$ while $\partial/\partial q_\mathcal{T}$ and $\partial/\partial p_\mathcal{T}$ act only on $\mathcal{T}_{M,n}(q,p)$. 

In many cases, the infinite series of differential operators is simply expressed in terms of the sine of the Poisson bracket differential operator, i.e., $(2/\hbar)\sin [(\hbar/2)(\partial/\partial q_H \,\partial/\partial p_T - \partial/\partial p_H \,\partial/\partial q_T )]$\cite{Jordan1961}. For our current purposes, we need to evaluate, or at least simplify, Eq. (\ref{moyalht}) for a given Hamiltonian. We substitute our assumed solution into Eq. (\ref{moyalht}) which leads to
\begin{equation}\label{moyal2}
1=\sum_{r=0}^{\infty}\,\sum_{n=0}^{\infty}\,\frac{\hbar^{2r+n}}{2^{2r}}\frac{(-1)^r}{(2r+1)!} \,\left[ \frac{\partial}{\partial q_H} \frac{\partial}{\partial p_\mathcal{T}}-\frac{\partial}{\partial p_H}\,\frac{\partial}{\partial q_\mathcal{T}}\right]^{2r+1}H(q,p) \,\mathcal{T}_{M,n}(q,p).
\end{equation}
This expansion can be simplified by recognizing the vanishing of the factors $\partial^{n+m}\,H(q,p)/\partial q^n\,\partial p^m=0$ for all $n,m \ge1$ with $H(q,p)=p^2/2\mu+V(q)$. Hence, Eq. (\ref{moyal2}) reads as
\begin{equation}\label{moyal3}
1=\sum_{r=0}^{\infty}\,\sum_{n=0}^{\infty}\,\frac{\hbar^{2r+n}}{2^{2r}}\frac{(-1)^r}{(2r+1)!} \,\left[ \frac{\partial^{2r+1}}{\partial q_H^{2r+1}}\, \frac{\partial^{2r+1}}{\partial p_\mathcal{T}^{2r+1}}-\frac{\partial^{2r+1}}{\partial p_H^{2r+1}}\,\frac{\partial^{2r+1}}{\partial q_\mathcal{T}^{2r+1}}\right]H(q,p) \,\mathcal{T}_{M,n}(q,p).
\end{equation}
Separating the even and odd terms in the summation along $n$, interchanging the order of summations between $r$ and $n$ using the identity
\begin{equation}\label{intersumm}
\sum_{n=0}^{\infty}\sum_{k=0}^{\infty}\,A(k,n)=\sum_{n=0}^{\infty}\sum_{k=0}^{n}\,A(k,n-k),
\end{equation}
and imposing uniqueness of power series with respect to $\hbar$, we arrive at the following set of equations,
\begin{equation}\label{pbclass}
1=\{H(q,p),\mathcal{T}_{M,0}(q,p)\}_{PB},
\end{equation}
\begin{equation}\label{evenhbar}
0=\sum_{n=1}^{\infty}\,\hbar^{2n}\,\sum_{r=0}^{2n}\,\frac{1}{2^{2r}}\frac{(-1)^r}{(2r+1)!} \,\left[ \frac{\partial^{2r+1}}{\partial q_H^{2r+1}}\, \frac{\partial^{2r+1}}{\partial p_\mathcal{T}^{2r+1}}-\frac{\partial^{2r+1}}{\partial p_H^{2r+1}}\,\frac{\partial^{2r+1}}{\partial q_\mathcal{T}^{2r+1}}\right]H(q,p) \,\mathcal{T}_{M,2n-2r}(q,p),
\end{equation}
\begin{equation}\label{oddhbar}
0=\sum_{n=0}^{\infty}\,\hbar^{2n+1}\,\sum_{r=0}^{2n+1}\,\frac{1}{2^{2r}}\frac{(-1)^r}{(2r+1)!} \,\left[ \frac{\partial^{2r+1}}{\partial q_H^{2r+1}}\, \frac{\partial^{2r+1}}{\partial p_\mathcal{T}^{2r+1}}-\frac{\partial^{2r+1}}{\partial p_H^{2r+1}}\,\frac{\partial^{2r+1}}{\partial q_\mathcal{T}^{2r+1}}\right]H (q,p)\,\mathcal{T}_{M,2n-2r+1}(q,p).
\end{equation}

One readily identifies Eq. (\ref{pbclass}) as the Poisson bracket between the system Hamiltonian and the classical arrival time, i.e., $\mathcal{T}_{M,0}(q,p)=\tau_{M,0}(q,p) \subset \mathcal{T}_C(q,p)$, which we have solved in Section (\ref{sec:PB}). Hence, $\mathcal{T}_{M,0}(q,p)$ is already known. Our immediate goal then is to determine the factors $\mathcal{T}_{M,2n}(q,p)$ and $\mathcal{T}_{M,2n+1}(q,p)$ from the partial differential equations set by Eqs. (\ref{evenhbar}) and (\ref{oddhbar}), respectively. 

\subsubsection{Odd powers of $\hbar$}

Let us solve $\mathcal{T}_{M,2n+1}(q,p)$ first. Imposing the vanishing of $\mathcal{T}_{M,n}(q,p)$ for negative $n$, the $r$-indexed sum in Eq. (\ref{oddhbar}) is nonzero only for $r\le n$. Isolating the $r=0$ term, shifting index from $r \to r-1$, and using the following simplifications
\begin{equation}\label{partialH}
\frac{\partial H(q,p)}{\partial q}=\frac{\partial V(q)}{\partial q}; \quad \quad \frac{\partial H(q,p)}{\partial p}=\frac{p}{\mu}; \quad\quad
\frac{\partial^{2r+1} H(q,p)}{\partial p^{2r+1}}=0; \quad r\ge 1,
\end{equation}
we arrive at the following partial differential equation
\begin{equation}\label{0eq2}
\frac{\partial \mathcal{T}_{M,2n+1}(q,p)}{\partial q}=\frac{\mu}{p}\frac{\partial V(q)}{\partial q}\frac{\partial \mathcal{T}_{M,2n+1}(q,p)}{\partial p}+\frac{\mu}{p}\sum_{r=0}^{2n}\,\left(\frac{1}{2}\right)^{2r}\frac{(-1)^r}{(2r+3)!} \,\frac{\partial^{2r+3} V(q)}{\partial q^{2r+3}}\frac{\partial^{2r+3} \, \mathcal{T}_{M,2n-2r-1}(q,p)}{\partial p^{2r+3}}.
\end{equation}

To solve the above recurrence relation, we need to identify the appropriate initial condition first. This is done by setting $n=0$ and using the boundary condition $\mathcal{T}_{M,-1}(q,p)=0$ in accordance to Eq. (\ref{bcnegn}). We then find
\begin{equation}\label{todd1}
\begin{split}
\frac{\partial \mathcal{T}_{M,1}(q,p)}{\partial q}=\frac{\mu}{p}\frac{\partial V(q)}{\partial q}\,\frac{\partial \mathcal{T}_{M,1}(q,p)}{\partial p}.
\end{split}
\end{equation}
Equation (\ref{todd1}) is a first-order partial differential equation with two possible general solutions. The first solution is obtained by the method of characteristics and is given by
\begin{equation}\label{gensol2}
\mathcal{T}_{M,1}(q,p)=F\left(\frac{p^2}{2\mu}+V(q)+C\right)=F\left(H(q,p)+C\right),
\end{equation}
where $F(H(q,p)+C)$ is any function of the Hamiltonian plus some arbitrary constant $C$. Note, however, that this solution is even in $p$ violating the time-reversal symmetry requirement (see Eq. (\ref{trversalsym})). Hence, it cannot be an acceptable solution. On the other hand, the other possible solution is simply $\mathcal{T}_{M,1}(q,p)=C$ for some constant $C$. Nonetheless, a non-zero $C$ solution is also not allowed for the same reason. We are then left with the trivial solution, $\mathcal{T}_{M,1}(q,p)=0$, which could accommodate both Eqs. (\ref{trversalsym}) and (\ref{todd1}). This solution ultimately suggests the vanishing of the term $\hbar \, \mathcal{T}_{M,1}(q,p)$ in Eq. (\ref{phasespacet}).

We repeat the same analysis for $n\ge1$, while imposing the boundary condition $\mathcal{T}_{M,-n}(q,p)=0$ and initial condition $\mathcal{T}_{M,1}(q,p)=0$. In general, we arrive at the following form
\begin{equation}\label{0eq3}
\frac{\partial \mathcal{T}_{M,2n+1}(q,p)}{\partial q}=\frac{\mu}{p}\frac{\partial V(q)}{\partial q}\,\frac{\partial \mathcal{T}_{M,2n+1}(q,p)}{\partial p}.
\end{equation}
Similar to Eq. (\ref{todd1}), time-reversal symmetry requirement only allows the solution
\begin{equation}
\mathcal{T}_{M,2n+1}(q,p)=0.
\end{equation}
Hence, odd powers of $\hbar$ do not contribute to $\mathcal{T}_M(q,p)$. We can then rewrite the Moyal arrival time $\mathcal{T}_M(q,p)$ as
\begin{equation}\label{phasespacet2}
\mathcal{T}_M(q,p)=\sum_{n=0}^{\infty}\hbar^{2n}\,\mathcal{T}_{M,2n}(q,p),
\end{equation}
showing only an even dependence on $\hbar$.

\subsubsection{Even powers of $\hbar$}

Our next goal is to solve for $\mathcal{T}_{M,2n}(q,p)$ from Eq. (\ref{evenhbar}). We impose uniqueness of power series, separate the $r=0$ term, and use the same simplifications defined by Eq. (\ref{partialH}).  We then determine the following partial differential equation 
\begin{equation}\label{t2n2n}
\frac{\partial \mathcal{T}_{M,2n}(q,p)}{\partial q}=\frac{\mu}{p}\frac{\partial V(q)}{\partial q}\,\frac{\partial \mathcal{T}_{M,2n}(q,p)}{\partial p}+\frac{\mu}{p}\sum_{r=1}^{2n}\,\left(\frac{1}{2}\right)^{2r}\frac{(-1)^r}{(2r+1)!} \,\frac{\partial^{2r+1} \, V(q)}{\partial q^{2r+1}}\, \frac{\partial^{2r+1} \, \mathcal{T}_{M,2(n-r)}(q,p)}{\partial p^{2r+1}},
\end{equation}
for all $n\ge1$. We convert the above differential equation into an integral equation by integrating both sides of Eq. (\ref{t2n2n}) along $q'$ from $0$ to $q$. We arrive at the integral equation
\begin{equation}\label{wwtn}
\mathcal{T}_{M,2n}(q,p)=\frac{\mu}{p}\,\sum_{r=0}^{2n}\,\frac{(-1)^r}{2^{2r}\,(2r+1)!} \,\int_{0}^{q}dq'\, \frac{\partial^{2r+1} \, V(q')}{\partial q'^{2r+1}}\, \frac{\partial^{2r+1} \, \mathcal{T}_{M,2(n-r)}(q',p) }{\partial p^{2r+1}}.
\end{equation}
Equation (\ref{wwtn}) can also be solved by the method of successive approximation, similar to the methods of Section (\ref{sec:PB}). Let our zeroth-order approximation be
\begin{equation}\label{zerothtn0}
F_{0}(q,p)=\frac{\mu}{p}\,\sum_{r=1}^{2n}\,\frac{(-1)^r}{2^{2r}\,(2r+1)!} \,\int_{0}^{q}dq'\, \frac{\partial^{2r+1} \, V(q')}{\partial q'^{2r+1}}\, \frac{\partial^{2r+1} \, \mathcal{T}_{M,2(n-r)}(q',p) }{\partial p^{2r+1}},
\end{equation}
which is independent of $\mathcal{T}_{M,2n}(q',p)$. The $m$th-order approximation is determined from the recurrence relation,
\begin{equation}\label{iteratetnm}
F_{m}(q,p)=F_{0}(q,p)+\frac{\mu}{p}\int_{0}^{q}dq' \, \frac{\partial V(q')}{\partial q'}\,\frac{\partial F_{m-1}(q',p)}{\partial p},
\end{equation}
for $m \ge 1$. Equation (\ref{wwtn}) then corresponds to Eq. (\ref{iteratetnm}) in the limit $m \to \infty$. Hence, the required solution $\mathcal{T}_{M,2n}(q,p)$ is determined from the relation $\mathcal{T}_{M,2n}(q,p)=\lim\limits_{m\to\infty}F_{m}(q,p).$

As shown in Appendix (\ref{appendix2}), 
we arrive at the following general form for arbitrary $m$,
\begin{equation}\label{taunm}
\begin{split}
F_{m}(q,p)=\mu\sum_{r=1}^{2n}\frac{(-1)^r}{2^{2r}\,(2r+1)!} \int_{0}^{q}dq'\sum_{j=0}^{m}\frac{\big(V(q)-V(q')\big)^j}{j!}\left(\frac{\mu}{p}\frac{\partial}{\partial p}\right)^j
\frac{1}{p} \frac{\partial^{2r+1} V(q')}{\partial q'^{2r+1}} \frac{\partial^{2r+1} \mathcal{T}_{M,2(n-r)}(q',p) }{\partial p^{2r+1}}.
\end{split}
\end{equation}
In the limit $m \to \infty$, we finally get the solution
\begin{equation}\label{tnqpfinalb}
\begin{split}
\mathcal{T}_{M,2n}(q,p)&=\mu\sum_{r=1}^{2n}\,\frac{(-1)^r}{2^{2r}\,(2r+1)!}\int_{0}^{q}dq'\mathrm{exp}\left[\big(V(q)-V(q')\big)\frac{\mu}{p}\frac{\partial}{\partial p}\right]\frac{1}{p} \frac{\partial^{2r+1} V(q')}{\partial q'^{2r+1}}\, \frac{\partial^{2r+1} \mathcal{T}_{M,2(n-r)}(q',p)}{\partial p^{2r+1}}.
\end{split}
\end{equation}
This equation is a recurrence equation whose initial condition is the local time of arrival $\mathcal{T}_{M,0}(q,p)=\tau_{M,0}(q,p)$. The $n$th term is known as long as the preceding terms are identified. 

Note that the right-hand side of Eq. (\ref{wwtn}) still depends on the factor $\mathcal{T}_{M,2n}(q,p)$, while the right-hand side of Eq. (\ref{tnqpfinalb}) is independent of $\mathcal{T}_{M,2n}(q,p)$. Therefore, the latter equation provides the desired solution for  $\mathcal{T}_{M,2n}(q,p)$. 

Defining $\tau_{M,n}(q,p)=\mathcal{T}_{M,2n}(q,p)$, the TOA phase-space function $\mathcal{T}_M(q,p)$ that satisfies the Moyal bracket relation with the system Hamiltonian formally appears as
\begin{equation}\label{toaexpandph}
\mathcal{T}_M(q,p)=\tau_{M,0}(q,p)+\sum_{n=1}^{\infty}\hbar^{2n}\,\tau_{M,n}(q,p),
\end{equation}
where each $\tau_{M,n}(q,p)$ can be obtained recursively from 
\begin{equation}\label{tnqpfinal}
\begin{split}
\tau_{M,n}(q,p)&=\mu\sum_{r=1}^{n}\,\frac{(-1)^r}{2^{2r}\,(2r+1)!} \,\int_{0}^{q}dq'\,\mathrm{exp}\left[\big(V(q)-V(q')\big)\frac{\mu}{p}\frac{\partial}{\partial p}\right]\frac{1}{p} \frac{\partial^{2r+1} \, V(q')}{\partial q'^{2r+1}}\, \frac{\partial^{2r+1} \, \tau_{M,n-r}(q',p)}{\partial p^{2r+1}}.
\end{split}
\end{equation}
For reference, the first two $\hbar-$dependent terms are given by
\begin{equation}\label{wwtn1}
\hbar^2\tau_{M,1}(q,p)=-\frac{\mu}{24}\hbar^2\,\int_{0}^{q}dq'\,\,\mathrm{exp}\left[\left(V(q)-V(q')\right)\,\frac{\mu}{p}\frac{\partial}{\partial\,p}\right]\frac{1}{p} \,\frac{\partial^{3} \, V(q')}{\partial q'^{3}}\, \frac{\partial^{3} \, \tau_{M,0}(q',p) }{\partial p^{3}},
\end{equation}

\begin{equation}\label{wwtn2}
\begin{split}
\hbar^4\tau_{M,2}(q,p)=&-\frac{\mu}{24}\hbar^4\,\int_{0}^{q}dq'\,\,\mathrm{exp}\left[\left(V(q)-V(q')\right)\,\frac{\mu}{p}\frac{\partial}{\partial\,p}\right]\frac{1}{p} \,\frac{\partial^{3} \, V(q')}{\partial q'^{3}}\, \frac{\partial^{3} \, \tau_{M,1}(q',p) }{\partial p^{3}}\\
&+\frac{\mu}{16\cdot5!}\hbar^4\,\int_{0}^{q}dq'\,\,\mathrm{exp}\left[\left(V(q)-V(q')\right)\,\frac{\mu}{p}\frac{\partial}{\partial\,p}\right]\frac{1}{p} \,\frac{\partial^{5} \, V(q')}{\partial q'^{5}}\, \frac{\partial^{5} \, \tau_{M,0}(q',p) }{\partial p^{5}},
\end{split}
\end{equation}
which can be simplified further for a given potential $V(q)$. 

Now, if the Moyal bracket can be interpreted as the formal deformation of the Poisson bracket, we interpret the quantum image $\mathcal{T}_M(q,p)$ as the Moyal deformation of the classical arrival time $\mathcal{T}_C(q,p)$, with $\tau_{M,0}(q,p) \subset \mathcal{T}_{C}(q,p)$. 

Notice that the terms $\hbar^{2n}\,\tau_{M,n}(q,p)$ vanishes for linear systems but generally non-vanishing for nonlinear systems. For linear systems of the form $V(q) = a + bq + cq^2$, the factor  $\partial^{2r+1} \, V(q')/\partial q'^{2r+1}$ is always zero for $r \ge 1$. Hence, there are no quantum corrections to the corresponding classical time of arrival so that $\mathcal{T}_M(q,p)=\mathcal{T}_C(q,p)$. For this case, both the classical and quantum observables coincide, and the dynamical algebras they satisfy are the same. This is expected since the Moyal bracket reduces to the Poisson bracket if either $H(q,p)$ or $\mathcal{T}_M(q,p)$ is of quadratic or of lower order in the $(q,p)$ variables. Within the quantum phase-space formulation, the classical and local arrival times are already sufficient for specific quantum arrival time problems involving a free particle, a particle in a gravitational field, and a particle in the presence of a harmonic oscillator, among others. The required algebra in accordance with Eq. (\ref{teccr}) is guaranteed.

On the other hand, the factors $\partial^{2r+1} \, V(q')/\partial q'^{2r+1}$ in Eq. (\ref{tnqpfinal}) are generally nonzero for nonlinear systems of the form $V(q)=\sum_{n=0}^{\infty} a_n q^n$, where at least the coefficient $a_3$ is non-vanishing. This makes sense since the Poisson and Moyal brackets no longer coincide. In the quantum TOA problem, obstructions to quantization \cite{Gotay1996,Gotay1999} arise if one ignores the $\hbar$-dependent terms in Eq. (\ref{toaexpandph}). These obstructions can then be bypassed by incorporating the $\hbar$-dependent terms not predicted by the theory of quantized TOA operators.

\subsubsection{Satisfying time-reversal symmetry}

Although we have solved for the Moyal arrival time $\mathcal{T}_M(q,p)$, we have yet to show that it satisfies the time-reversal symmetry. For this case, showing its dependence on odd powers of $p$ suffices.

For $n=1$, we expand Eq. (\ref{wwtn1}) and perform the indicated partial derivatives involving the LTOA with respect to $p$. The evaluation can be facilitated by rewriting $\tau_{M,0}(q,p)$ as $\tau_{M,0}(q,p)=-\sum_{k=0}^{\infty}Q(q)p^{-2k-1}$ for some function $Q(q)$ which is not important for our current purposes. Equation (\ref{wwtn1}) then leads to
\begin{equation}\label{ht1pdep}
\begin{split}
\hbar^2\tau_{M,1}(q,p)&=\frac{\mu}{24}\hbar^2\,\int_{0}^{q}dq'\,\sum_{k=0}^{\infty}Q(q')\,\,\frac{\partial^{3} \, V(q')}{\partial q'^{3}}\,\mathrm{exp}\left[\left(V(q)-V(q')\right)\,\frac{\mu}{p}\frac{\partial}{\partial\,p}\right]\frac{1}{p}\, \frac{\partial^{3} \, p^{-2k-1}}{\partial p^{3}}\\
&=\frac{\mu}{24}\hbar^2\,\int_{0}^{q}dq'\,\sum_{k=0}^{\infty}Q(q')\,\,\frac{\partial^{3} \, V(q')}{\partial q'^{3}}\,\sum_{l=0}^{\infty}\frac{\mu^l}{l!}\left(V(q)-V(q')\right)^l\,\left(\frac{1}{p}\frac{\partial}{\partial\,p}\right)^{l+1}\, \frac{\partial^{2} \, p^{-2k-1}}{\partial p^{2}}\\
&=\hbar^2\,\int_{0}^{q}dq'\,\sum_{k,l=0}^{\infty}Q^{(1)}_{k,l}(q')\,p^{-2k-2l-5},
\end{split}
\end{equation}
where we have written all $q'$-dependent factors as $Q^{(1)}_{k,l}(q')$ for convenience. The last line already shows an odd $p$-dependence. For some appropriate neighborhood, the term $\hbar^2 \tau_{M,1}(q,p)$ converges to $p^{-5}\,F_1\left(q,p^2\right)$ 
where $F_1\left(q,p^2\right)$ is a particular function dependent on the even powers of $p$. Therefore, $\hbar^2\tau_{M,1}(q,p)$ is odd in $p$. 

For $n=2$, Eqs. (\ref{tnqpfinal}) and (\ref{ht1pdep}) also lead to
\begin{equation}\label{ht2pdep}
\begin{split}
\hbar^4\tau_{M,2}(q,p)=\int_{0}^{q}dq'\,\sum_{k,l=0}^{\infty}Q^{(1)}_{k,l}(q')\,p^{-2k-2l-9},
\end{split}
\end{equation}
so that the third term of $\mathcal{T}_M(q,p)$ converges to $\hbar^4 \tau_2(q,p)=p^{-9}\,F_2\left(q,p^2\right)$. For arbitrary $n$, one can show that
\begin{equation}\label{htnpdep}
\begin{split}
\hbar^{2n}\tau_{M,n}(q,p)=\hbar^{2n}\,\int_{0}^{q}dq'\,\sum_{k,l=0}^{\infty}Q^{(n)}_{k,l}(q')\,p^{-2k-2l-4n-1},
\end{split}
\end{equation}
which clearly illustrates the odd-dependence of $\hbar^{2n}\tau_{M,n}(q,p)$ on $p$. Hence, $\mathcal{T}_M(q,p)$ satisfies the condition $\mathcal{T}_M(q,p)=-\mathcal{T}_M(q,-p).$

We have now satisfied our first objective and demonstrated how one obtains the quantum image of the classsical arrival time using the quantum phase space formulation. Of course, our approach is not limited to arrival times and can be extended to other observables satisfying specific conjugacy relations. 

\section{Weyl mapping of the Moyal time of arrival function}\label{sec:weylmap}

We now determine the TOA operator corresponding to $\mathcal{T}_M(q,p)$. We call this operator the Moyal TOA operator $\hat{\mathrm{T}}_M$ to differentiate it initially from the supraquantized and Weyl-quantized TOA operators, $\hat{\mathrm{T}}_S$ and $\hat{\mathrm{T}}_W$, respectively.

In the standard formulation of quantum phase space, observables are usually represented by smooth functions and its operator correspondence is defined in the usual Hilbert space setting of quantum mechanics. In the quantum TOA problem, however, $\mathcal{T}_M(q,p)$ is singular in $p$. Nevertheless, the existence and characterization of $\mathcal{T}_M(q,p)$ as a phase space observable follows from $\mathcal{T}_C(q,p)$. In fact, we argue that the divergence in $p$ is necessary since a particle with zero momentum does not arrive at the desired arrival point. Besides, the singularity along $p=0$ poses no problem as long as we manipulate $\mathcal{T}_M(q,p)$ in the distributional sense.

The proper operator equivalence of our phase space approach is then defined within the rigged Hilbert space (RHS) formulation of quantum theory. It is basically a Hilbert space endowed with the theory of distributions \cite{Madrid2005, Madrid2002}. Formally, the RHS offers a proper treatment of observables that are generally unbounded so that their spectrum have in general continuous part. These observables then have eigenfunctions that are non-normalizable and non-square integrable, thus are clearly outside of the usual Hilbert space of quantum mechanics. These singular objects, however, can be treated as distributions in RHS \cite{Pablico2023}, which is ideal to our current purposes. 

We now consider a structureless particle in the real line described by the Hilbert space $\mathcal{H}=L^2(\mathcal{R})$. We can choose a rigging defined by $\Phi^\times \supset L^2(\mathcal{R}) \supset \Phi $, where $\Phi$ is the fundamental space of infinitely differentiable functions in the real line with compact supports and $\Phi^\times$ is the space of functionals on $\Phi$. The standard Hilbert space formulation of quantum mechanics is obtained by taking the closures in $\Phi$ with respect to the metric $L^2(\mathcal{R})$ \cite{Galapon2004,Galapon2018}. In this framework, our arrival time observable $\mathcal{T}_M(q,p)$ appears as the mapping $\hat{\mathrm{T}}_M: \Phi  \to \Phi^\times$ and is formally defined by the following integral operator
\begin{equation}\label{operator}
(\hat{\mathrm{T}}_M\,\varphi )(q)=\int_{-\infty}^{\infty} \,dq' \,\langle q|\hat{\mathrm{T}}_M|q'\rangle \,\varphi (q'),
\end{equation}
in coordinate representation. Our goal  then is to determine the appropriate kernel $\langle q|\hat{\mathrm{T}}_M|q'\rangle$ that corresponds to $\mathcal{T}_M(q,p)$.

\subsection{Weyl map of the boundary conditions}\label{subsec:BC}

Following Weyl's prescription \cite{Zachos2001,Zachos2005}, the kernel $\langle q|\hat{\mathrm{T}}_M|q'\rangle$ can be obtained from $\mathcal{T}_M(q,p)$ through the mapping
\begin{equation}\label{inversefor}
\langle q|\hat{\mathrm{T}}_M|q'\rangle =\frac{1}{2\pi\hbar}\, \int_{-\infty}^{\infty}dp\,\, \mathcal{T}_M\left(\frac{q+q'}{2},p\right)\mathrm{exp}\left[\frac{i}{\hbar}(q-q')p\right].
\end{equation}
Meanwhile, the Moyal arrival time $\mathcal{T}_M(q,p)$ can be obtained from the inverse transform of Eq. (\ref{inversefor}), 
\begin{equation}\label{oinverse}
\mathcal{T}_M(q,p)=\int_{-\infty}^{\infty}d\nu\, \left\langle q+\frac{\nu}{2}\left|\hat{\mathrm{T}}_M\right|q-\frac{\nu}{2}\right\rangle \mathrm{e}^{-i \nu p/\hbar}. 
\end{equation}
The last two relations provide a one-to-one correspondence between RHS and quantum phase-space observables. 

The form of $\mathcal{T}_M(q,p)$ expressed as an infinite series in powers of $\hbar^2$ suggests the following kernel of the Moyal TOA operator
\begin{equation}\label{kernelinfinite}
\langle q|\hat{\mathrm{T}}_{M}|q'\rangle = \sum_{n=0}^\infty\,\langle q|\hat{\mathrm{T}}_{M,n}|q'\rangle,
\end{equation}
where each $\langle q|\hat{\mathrm{T}}_{M,n}|q'\rangle$ is the Weyl map of $\hbar^{2n}\tau_{M,n}(q,p)$ in accordance to Eq. (\ref{inversefor}).

We begin by determining the Weyl map of the boundary conditions we have imposed on $\mathcal{T}_M(q,p)$. This is done to identify the specific conditions satisfied by $\langle q|\hat{\mathrm{T}}_M|q'\rangle$ and compare it with the conditions satisfied by the kernel $\langle q|\hat{\mathrm{T}}_S|q'\rangle$ of the supraquantized TOA operator. Taking the complex conjugate of Eq. (\ref{inversefor}) and imposing hermiticity given by $\mathcal{T}_M(q,p)=\mathcal{T}_M(q,p)^*$, we arrive at
\begin{equation}\label{tkcon1}
\langle q|\hat{\mathrm{T}}_M|q'\rangle=\langle q'|\hat{\mathrm{T}}_M|q\rangle^*. 
\end{equation}
On the other hand, changing variables from $p$ to $-p$ in Eq. (\ref{inversefor}) and imposing time-reversal symmetry, $\mathcal{T}_M(q,p)=-\mathcal{T}_M(q,-p)$, we also determine the following condition
\begin{equation}\label{tkcon2}
\langle q|\hat{\mathrm{T}}_M|q'\rangle=-\langle q|\hat{\mathrm{T}}_M|q'\rangle^*.
\end{equation}

In addition, we should be able to recover the known expression for the free particle case in the limit of vanishing potential. Direct substitution of  $\mathcal{T}_F(q,p)=-\mu q/p$ to Eq. (\ref{inversefor}) gives the kernel
\begin{equation}\label{inverseforfree}
\langle q|\hat{\mathrm{T}}_F|q'\rangle =-\frac{\mu}{2\pi\hbar}\,(q+q') \int_{-\infty}^{\infty}dp\,p^{-1}\,\mathrm{exp}\left[\frac{i}{\hbar}(q-q')p\right].
\end{equation}
The integral along $p$ is easily evaluated using the following distributional identity  \cite{Gelfand1964}
\begin{equation}\label{disiden}
\int_{-\infty}^{\infty} x^{-m}\, e^{ix\nu} dx =\pi \frac{i^m\, \nu^{m-1}\,\mathrm{sgn}(\nu)}{\,(m-1)!}.
\end{equation}
Hence, Eq. (\ref{inverseforfree}) simplifies to
\begin{equation}\label{freetkernel}
\langle q|\hat{\mathrm{T}}_F|q'\rangle=\frac{\mu}{i\hbar} \, \mathrm{sgn}(q-q')\, \left(\frac{q+q'}{4}\right),
\end{equation}
which gives the time kernel of the TOA operator $\hat{\mathrm{T}}_F$ following Eq. (\ref{operator}). The Moyal kernel $\langle q|\hat{\mathrm{T}}_M|q'\rangle$ should then satisfy the relation
\begin{equation}
\lim_{V(q) \to 0} \langle q|\hat{\mathrm{T}}_M|q'\rangle \to \langle q|\hat{\mathrm{T}}_F|q'\rangle.
\end{equation}

The above conditions can already help us deduce the general form of $\langle q|\hat{\mathrm{T}}_M|q'\rangle$ for arbitrary potentials. Recall that time-reversal symmetry requires  $\mathcal{T}_M(q,p)$ to be odd in $p$. Hence, we can write $\mathcal{T}_M(q,p)=-\sum_{n=0}^{\infty}\hbar^{2n} F_n (q) \, p^{-2n-1}$ for some function $F_n (q)$. Equation (\ref{inversefor}) then leads to
\begin{equation}\label{inversefortq}
\langle q|\hat{\mathrm{T}}_M|q'\rangle =-\frac{\mu}{2\pi\hbar}\,\sum_{n=0}^{\infty}\,F_n(q,q') \int_{-\infty}^{\infty}dp\,\,p^{-2k-1}\,\mathrm{exp}\left[\frac{i}{\hbar}(q-q')p\right].
\end{equation}
Again, we use the same distributional identity defined by Eq. (\ref{disiden}) to get the following general form
\begin{equation}\label{kernelfactgen}
\langle q|\hat{\mathrm{T}}_M|q'\rangle =\frac{\mu}{i\hbar}\,\mathrm{sgn}(q-q')\,T_M(q,q'),
\end{equation}
for some kernel factor $T_M(q,q')$. For the free particle case, Eq. (\ref{freetkernel}) already suggests the time kernel factor $T_F(q,q')=(q+q')/4$ so that $T_M(q,q') \to T_F(q,q')$ in the limit $V(q) \to 0$.

Comparing Eqs. (\ref{kernelinfinite}) and (\ref{kernelfactgen}), we can then rewrite the kernel factor $T_M(q,q')$ as the expansion
\begin{equation}
T_M(q,q')=\sum_{n=0}^{\infty}T_{M,n}(q,q').
\end{equation}
Furthermore, imposing Eqs. (\ref{tkcon1}) and (\ref{tkcon2}) gives the following conditions on $T_M(q,q')$,
\begin{equation}
T_M(q,q')=T_M(q,q')^*\,\,; \,\,T_M(q,q')=T_M(q',q), 
\end{equation}
ensuring that hermiticity and time-reversal symmetry are satisfied. Finally, we can use the above conditions, coupled with the free-particle time kernel factor $T_F(q,q')$ to obtain the following additional boundary conditions on $T_M(q,q')$,
\begin{equation}
T_M(q,q)=\frac{q}{2} \,\,;\,\,T_M(q,-q)=0. 
\end{equation}

Interestingly, all conditions derived for $\langle q|\hat{\mathrm{T}}_M|q'\rangle$ and $T_M(q,q')$ are exactly the same boundary conditions for the kernel $\langle q|\hat{\mathrm{T}}_S|q'\rangle$ and $T_S(q,q')$, respectively, in the theory of supraquantization. In the next section, we determine explicitly the functional form of $T_M(q,q')$ from the Weyl map of $\mathcal{T}_M(q,p)$. 

\subsection{Weyl map of the LTOA}
We now quantize the LTOA given by Eq. (\ref{ltoapb}). Following Eq. (\ref{inversefor}), we find the following kernel
\begin{equation}\label{kernelto}
\begin{split}
\langle q|\hat{\mathrm{T}}_{M,0}|q'\rangle =-\frac{\mu}{\hbar}\,\int_{0}^{\frac{q+q'}{2}}ds \,\sum_{k=0}^{\infty}(-1)^k \frac{(2k-1)!!}{k!} \mu^{k} \left(V\left(\frac{q+q'}{2}\right)-V(s)\right)^k\int_{-\infty}^{\infty}dp\,\frac{\mathrm{e}^{\frac{i}{\hbar}(q-q')p}}{ p^{2k+1}}.
\end{split}
\end{equation}
The integral along $p$ can be evaluated using Eq. (\ref{disiden}) in the distributional sense. Hence, we arrive at
\begin{equation}\label{kernelto1}
\langle q|\hat{\mathrm{T}}_{M,0}|q'\rangle=\frac{\mu}{i\hbar}\,\mathrm{sgn}(q-q')\,\,\frac{1}{4}\int_{0}^{q+q'}ds \,\sum_{k=0}^{\infty}\left(\frac{\mu}{2\hbar^2}\right)^k \frac{(q-q')^{2k}}{k!\,k!} \left(V\left(\frac{q+q'}{2}\right)-V\left(\frac{s}{2}\right)\right)^k.
\end{equation}
The sum along $k$ can be closed as a specific hypergeometric function defined by
\begin{equation}\label{hypergeom}
{}_0F_{1}(;b;z)=\sum_{k=0}^{\infty}\frac{z^k}{(b)_k k!}.
\end{equation}
We then arrive at the kernel $\langle q|\hat{\mathrm{T}}_{M,0}|q'\rangle =(\mu/i\hbar)\,\mathrm{sgn}(q-q')\,T_{M,0}(q,q'),$
where we identify $T_{M,0}(q,q')$ as
\begin{equation}\label{t0qqp}
T_{M,0}(q,q')=\frac{1}{4}\int_{0}^{q+q'} ds \, {}_0F_{1}\left(;1;\left(\frac{\mu}{2\hbar^2}\right)(q-q')^2\left[V\left(\frac{q+q'}{2}\right)-V\left(\frac{s}{2}\right)\right]\right).
\end{equation}
Indeed, $\langle q|\hat{\mathrm{T}}_{M,0}|q'\rangle$ is of the form predicted by Eq. (\ref{kernelfactgen}) and correctly satisfies the conditions set by Eqs. (\ref{tkcon1}) and (\ref{tkcon2}).

Equation (\ref{t0qqp}) corresponds to the time kernel factor of the Weyl-quantized TOA operator $\hat{\mathrm{T}}_W$, e.g., $T_{W}(q,q')=T_{M,0}(q,q')$ introduced in Ref. \cite{Galapon2001}  and has been applied to several TOA problems in the literature (see for instance \cite{Galapon2001,Galapon2004,Galapon2006,Galapon2009,Galapon2009a,Galapon2009,Galapon2012,Sombillo2014,Sombillo2016,Galapon2018,Sombillo2018,Pablico2020}). It also coincides with the leading time kernel factor of the supraquantized TOA operator $T_{S,0}(q,q')$. Hence, we arrive at the relation $T_{W}(q,q')=T_{M,0}(q,q')=T_{S,0}(q,q')$.

\subsection{Weyl map of the quantum corrections}

Next is to determine the Weyl map of the terms $\hbar^{2n}\tau_{M,n}(q,p)$. We do so by iteration. We start from obtaining the Weyl maps of $\hbar^{2}\tau_{M,1}(q,p)$, and $\hbar^{4}\tau_{M,2}(q,p)$, to get $\langle q|\hat{\mathrm{T}}_{M,1} |q'\rangle$, and $\langle q|\hat{\mathrm{T}}_{M,2} |q'\rangle$, respectively. We then infer a general form for $\langle q|\hat{\mathrm{T}}_{M,n} |q'\rangle$ for arbitrary $n$. The inferred solution will be validated formally by mathematical induction. Comparison with the results of supraquantization will then be made. 

For $n=1$, substitution of Eq. (\ref{wwtn1}) into Eq. (\ref{inversefor}) leads to
\begin{equation}\label{qt1q'}
\langle q|\hat{\mathrm{T}}_{M,1}|q'\rangle =-\frac{\mu \hbar}{48\pi} \sum_{l=0}^{\infty}\,\frac{\mu^l}{l!}\,\int_{0}^{\frac{q+q'}{2}}ds\, \left[V\left(\frac{q+q'}{2}\right)-V(s)\right]^l \,\frac{\partial^3 V(q')}{\partial q'^3}\,J(q,q',s),
\end{equation}
where the factor $J(q,q',s)$ involves a distributional integral along $p$ defined by
\begin{equation}\label{jqqs}
J(q,q',s)=\int_{-\infty}^{\infty}dp\,\mathrm{e}^{\frac{i}{\hbar}(q-q')p} \left(\frac{1}{p}\frac{\partial}{\partial p}\right)^l \,\frac{1}{p}\,\frac{\partial^3 \tau_0(s,p)}{\partial p^3}.
\end{equation}
Performing the indicated partial derivatives involving the LTOA, one arrives at the following result
\begin{equation}
\left(\frac{1}{p}\frac{\partial}{\partial p}\right)^l \,\frac{1}{p}\,\frac{\partial^3 \tau_0(s,p)}{\partial p^3}=\sum_{k=0}^{\infty}(-1)^{k+l} \frac{(2k+2)(2k+2l+3)!!}{k!} \frac{\mu^{k}}{p^{2k+2l+5}}\int_{0}^{s}ds' \, \left(V(s)-V(s')\right)^k.
\end{equation}
Thus, we can rewrite Eq. (\ref{jqqs}) as
\begin{equation}
J(q,q',s)=\sum_{k=0}^{\infty}(-1)^{k+l} \frac{(2k+2)(2k+2l+3)!!}{k!}\mu^k\int_{0}^{s}ds' \, \left(V(s)-V(s')\right)^k\,\int_{-\infty}^{\infty}dp\,\mathrm{e}^{\frac{i}{\hbar}(q-q')p}\,p^{-2k-2l-5}.
\end{equation}
The integral along $p$ can now be evaluated using the same distributional integral given by Eq. (\ref{disiden}) so that we arrive at
\begin{equation}\label{jqqsb}
J(q,q',s)=\frac{\pi i }{4\hbar^4}\,\mathrm{sgn}(q-q')\,\sum_{k=0}^{\infty} \frac{(2k+2)}{k!}\frac{\mu^k}{(2\hbar^2)^{k+l}}\frac{(q-q')^{2k+2l+4}}{(k+l+2)!}\int_{0}^{s}ds' \, \left(V(s)-V(s')\right)^k.
\end{equation}

Substitution of $J(q,q',s)$ into Eq. (\ref{qt1q'}) results into a coupled double sum. They can be decoupled by rewriting the factor $(q-q')^{2k+2l+4}/(k+l+2)!$ in Eq. (\ref{jqqsb}) as an integral of the form given by
\begin{equation}\label{rewfrac}
\frac{(q-q')^{2k+2l+4}}{(k+l+2)!}=\frac{2}{k!\,l!\,(k+1)}\int_{0}^{q-q'}dw\,w^{2k+3}\,\left((q-q')^2-w^2\right)^l.
\end{equation}
Upon substituting Eqs. (\ref{jqqsb}) and (\ref{rewfrac}) in Eq. (\ref{qt1q'}), we now arrive at
\begin{equation}\label{qt1q'2}
\begin{split}
\langle q|\hat{\mathrm{T}}_{M,1}|q'\rangle =&\frac{\mu}{i\hbar}\mathrm{sgn}(q-q')\,\frac{\mu}{48 \hbar^2}\,\int_{0}^{q+q'}ds\,\,\frac{\partial^3 V(s/2)}{\partial s^3}\,\int_{0}^{q-q'}dw\,w^{3}\,\bigg\{\frac{1}{4}\,\int_{0}^{s}ds'\sum_{k=0}^{\infty}\left(\frac{\mu}{2\hbar^2}\right)^k \frac{w^{2k}}{k!\,k!}\\ &\times\left(V\left(\frac{s}{2}\right)-V\left(\frac{s'}{2}\right)\right)^k\bigg\}\,\,\left[
\sum_{l=0}^{\infty}\left(\frac{\mu}{2\hbar^2}\right)^k \frac{\left((q-q')-w^2\right)^l}{l!\,l!}\left(V\left(\frac{q+q'}{2}\right)-V\left(\frac{s}{2}\right)\right)^l\right].
\end{split}
\end{equation}

Notice that the two infinite series can now be closed as two specific hypergeometric functions (see Eq. (\ref{hypergeom})). Specifically, the factor inside the curly brackets is identified as the leading time kernel factor $T_{M,0}(s,w)$ in $(s,w)$ coordinates, in accordance to Eq. (\ref{t0qqp}). On the other hand, the factor inside the square brackets is exactly the expansion of Eq. (\ref{gsw}). Hence, the Weyl map of the term $\hbar^2\tau_{M,1}\,(q,p)$ appears as
$\langle q|\hat{\mathrm{T}}_{M,1} |q'\rangle =(\mu/i\hbar)\,\mathrm{sgn}(q-q')\,T_{M,1}(q,q')$ where
\begin{equation}\label{t10uvfinb}
\begin{split}
T_{M,1}(q,q')&=\frac{\mu}{48\hbar^2}\int_{0}^{q+q'} ds \, \frac{\partial^3 V(s/2)}{\partial s^3}\, \int_{0}^{q-q'} dw \, w^3 \, T_{M,0}(s,w)\,\,G(s,w),
\end{split}
\end{equation}
with $G(s,w)$ defined by Eq. (\ref{gsw}). Since $\hbar^{2}\tau_{M,1}(q,p)$ is interpreted as the leading quantum correction to the local arrival time, $\tau_{M,0}(q,p)$, the kernel factor $T_{M,1}(q,q')$ can be interpreted too as the leading correction to the Weyl-quantized time kernel factor $T_{W}(q,q')=T_{M,0}(q,q')$.

Following similar calculations, the Weyl map of $\hbar^4\tau_{M,2}(q,p)$ given by Eq. (\ref{wwtn2}) assumes a similar form given by $\langle q|\hat{\mathrm{T}}_{M,2} |q'\rangle =(\mu/i\hbar)\,\mathrm{sgn}(q-q')\,T_{M,2}(q,q')$ 
with the kernel factor 
\begin{equation}\label{t20uv}
\begin{split}
T_{M,2}(q,q')&=\frac{1}{4 \cdot 3!}\left(\frac{\mu}{2\hbar^2}\right)\int_{0}^{q+q'} ds \, \frac{\partial^3 V(s/2)}{\partial s^3} \int_{0}^{q-q'} dw \, w^3 \, T_{M,1}(s,w) \, G(s,w)\\
&+\frac{1}{16 \cdot 5!}\left(\frac{\mu}{2\hbar^2}\right)\int_{0}^{q+q'} ds \, \frac{\partial^5 V(s/2)}{\partial s^5} \int_{0}^{q-q'} dw \, w^5 \, T_{M,0}(s,w) \, G(s,w).
	\end{split}
\end{equation}
Note that the explicit forms of the kernels $\langle q|\hat{\mathrm{T}}_{M,1} |q'\rangle$ and $\langle q|\hat{\mathrm{T}}_{M,2} |q'\rangle$ allow us infer the following time kernel for $n \ge 1$ 
\begin{equation}\label{tnuvkernrl}
\langle q|\hat{\mathrm{T}}_{M,n}|q'\rangle =\frac{\mu}{i\hbar}\,\mathrm{sgn}(q-q')\,T_{M,n}(q,q'),
\end{equation}
where the $n$th kernel factor $T_{M,n}(q,q')$ assumes the form
\begin{equation}\label{tnuvfin}
T_{M,n}(q,q')=\left(\frac{\mu}{2\hbar^2}\right)\sum_{r=1}^{n} \frac{1}{(2r+1)!}\frac{1}{2^{2r}}\int_{0}^{q+q'} ds \, \frac{\partial s^{2r+1}}{\partial s^{2r+1}}V\left(\frac{s}{2}\right) \int_{0}^{q-q'} dw \, w^{2r+1} \, T_{M,n-r}(s,w)\, G(s,w).
\end{equation}
The leading kernel factor $T_{M,0}(q,q')$ serves as the initial condition of Eq. (\ref{tnuvfin}). The convergence of the above result is guaranteed by the continuity of the potential $V(q)$ and the absolute convergence of the hypergeometric function ${}_{p}F_{q}(a;b;z)$ for $p<q$. 

It is easy to check that Eqs. (\ref{tnuvkernrl}) and (\ref{tnuvfin}) satisfy the required hermiticity and time-reversal symmetry conditions discussed in Sec. (\ref{subsec:BC}). Additionally, setting $n=1$ and $n=2$, we obtain correctly Eqs. (\ref{t10uvfinb}) and (\ref{t20uv}), respectively. The above result is formally proven by mathematical induction, as shown in Appendix (\ref{app:tkfn}). That is, we assume $T_{M,n}(q,q')$ to be valid for some $n=k$, the next iterate $n=k+1$ also holds true. 

Now, what we have obtained is exactly Eq. (\ref{tnuvintro}), the unique solution of the time kernel equation predicted by the theory of supraquantization. Thus, we have the equalities $T_{M,n}(q,q') = T_{S,n}(q,q')$ and $\langle q|\hat{\mathrm{T}}_{M}|q'\rangle=\langle q|\hat{\mathrm{T}}_{S}|q'\rangle$. Our results imply that the solution of the TKE is simply the Weyl map of the Moyal arrival time $\mathcal{T}_M(q,p)$ obtained from the Moyal bracket relation with the system Hamiltonian. We can then conclude that the Moyal and supraquantized TOA operators are the same,
\begin{equation}
\hat{\mathrm{T}}_M = \hat{\mathrm{T}}_S. 
\end{equation}
This further suggests the equality of the theoretical predictions obtained from $\mathcal{T}_M(q,p)$ using phase space techniques and $\hat{\mathrm{T}}_S$ using RHS operator techniques, satisfying our second objective. Notice that even in the phase space formulation of quantum theory, the solution of the quantum TOA problem, especially for nonlinear systems, is nontrivial due to the non-commutative algebra satisfied by the Hamiltonian and the arrival time observable. 

Now, the leading term of the algebra-preserving TOA operator is simply the Weyl-quantized TOA operator. The succeeding terms of $\hat{\mathrm{T}}_M$ cannot be accounted for by the theory of quantized TOA operators but are properly accounted for in the current approach. It is now clear what we need to modify in the theory of quantized TOA operators. First, the time-energy conjugacy relation should be represented by the Moyal bracket, not the Poisson bracket, in phase space. Second, the Weyl quantization of $\mathcal{T}_C(q,p)$ must be replaced by the Weyl mapping of $\mathcal{T}_M(q,p)$. Hence, our third objective is satisfied. 

\subsection{An alternative derivation of the time kernel equation from phase space}

For completeness, we derive explicitly the time kernel equation from the Weyl map of $\mathcal{T}_M(q,p)$. Using Eqs. (\ref{toaexpandph}), (\ref{inversefor}) and (\ref{kernelfactgen}), we find
\begin{equation}\label{sgntqq}
\frac{\mu}{i\hbar}\,\mathrm{sgn}(q-q')\,T_M(q,q') =\sum_{n=0}^{\infty} \frac{\hbar^{2n}}{2\pi\hbar}\, \int_{-\infty}^{\infty}dp\,\, \mathcal{\tau}_{M,n}\left(\frac{q+q'}{2},p\right)\mathrm{exp}\left[\frac{i}{\hbar}(q-q')p\right]. 
\end{equation}
We isolate the leading term, $n=0$, and rewrite it as $(\mu/i\hbar)\,\mathrm{sgn}(q-q')\,T_{M,0}(q,q')$. Substitution of Eq. (\ref{wwtn}) into Eq. (\ref{sgntqq}) followed by a change of variables $u=q+q'$ and $v=q-q'$, we find 
\begin{equation}
\begin{split}
\frac{\mathrm{sgn}(v)}{i\hbar}T_M(u,v)=\frac{\mathrm{sgn}(v)}{i\hbar}T_{M,0}(u,v)\,+\,&\sum_{n=1}^{\infty} \frac{\hbar^{2n}}{2\pi\hbar}\sum_{r=0}^{n}\frac{(-1)^{r+1}}{2^{2r}(2r+1)!} \int_{0}^{\frac{u}{2}}ds\,\, \frac{\partial^{2r+1}  V(s)}{\partial s^{2r+1}}\\
&\times\int_{-\infty}^{\infty}dp\,\frac{e^{ivp/\hbar}}{p}\frac{\partial^{2r+1}  \tau_{M,n-r}(s,p) }{\partial p^{2r+1}}.
\end{split}
\end{equation}

We take the partial derivative with respect to $v$ of both sides of the above equation, impose the known boundary conditions $T_M(u,0)=u/4$ and $T_M(0,v)=0$, and perform integration by parts $2r+1$ times in the integral along $p$. We arrive at
\begin{equation}\label{tmuvbbb}
\begin{split}
\mathrm{sgn}(v)\frac{\partial T_M(u,v)}{\partial v} =\mathrm{sgn}(v)\frac{\partial T_{M,0}(u,v)}{\partial v}\,+\,&\sum_{n=1}^{\infty} \frac{\hbar^{2n}}{2\pi\hbar}\sum_{r=0}^{n}\frac{(-1)^{r+1}}{2^{2r}(2r+1)!}\,\int_{0}^{\frac{u}{2}}ds\,\, \frac{\partial^{2r+1}  V(s)}{\partial s^{2r+1}}\left(\frac{iv}{\hbar}\right)^{2r+1}\\
&\times\int_{-\infty}^{\infty}dp\,e^{ivp/\hbar}\,\tau_{M,n-r}(s,p).
\end{split}
\end{equation}
Note that the integral along $p$ can be rewritten as
\begin{equation}
\frac{\hbar^{2n}}{2\pi\hbar}\int_{-\infty}^{\infty}dp\,e^{ivp/\hbar}\,\tau_{n-r}(s,p) = \frac{\mu}{i\hbar}\mathrm{sgn}(v)\,T_{M,n-r}(s,v),
\end{equation}
which results from Eqs. (\ref{inversefor}) and (\ref{kernelfactgen}). Taking the partial derivative with respect to $u$ of both sides of Eq. (\ref{tmuvbbb}), one finds
\begin{equation}\label{tkesoll}
\frac{\partial^2 T_M(u,v)}{\partial u\partial v}=\frac{\partial^2 T_{M,0}(u,v)}{\partial u\partial v}+\frac{\mu}{2\hbar^2}\sum_{n=1}^{\infty}\sum_{r=0}^{n}\frac{v^{2r+1}}{2^{2r}(2r+1)!}  \frac{\partial^{2r+1}  V(u/2)}{\partial ^{2r+1} u}\,T_{M,n-r}(u,v).
\end{equation}
Using Eq. (\ref{t0qqp}), the leading term involving $\partial^2 T_0(u,v)/\partial u\partial v$ evaluates to
\begin{equation}
\frac{\partial^2 T_0(u,v)}{\partial u\partial v}=\frac{\mu}{2\hbar^2}v\  \frac{\partial  V(u/2)}{\partial u}\, T_0(u,v).
\end{equation}
Hence, we can rewrite Eq. (\ref{tkesoll}) as follows
\begin{equation}\label{tkesoll2b}
\frac{\partial^2 T_M(u,v)}{\partial u\partial v}=\frac{\mu}{2\hbar^2}\sum_{n=0}^{\infty}\sum_{r=0}^{n}\frac{v^{2r+1}}{2^{2r}(2r+1)!}  \frac{\partial^{2r+1}  V(u/2)}{\partial ^{2r+1}u} \,T_{M,n-r}(u,v).
\end{equation}

We can simplify the above equation by decoupling the two infinite series using the identity
\begin{equation}
\sum_{n=0}^{\infty}\sum_{r=0}^{n}B(r,n)=\sum_{n=0}^{\infty}\sum_{r=0}^{\infty}B(r,n+r),
\end{equation}
and keeping in mind that the complete form the Moyal kernel factor is $T_M(u,v)=\sum_{n=0}^{\infty}T_{M,n}(u,v)$. Hence, we find
\begin{equation}\label{tkesoll2}
\frac{\partial^2 T_M(u,v)}{\partial u\partial v}=\frac{\mu}{2\hbar^2}\sum_{r=0}^{\infty}\frac{v^{2r+1}}{2^{2r}(2r+1)!}  \frac{\partial^{2r+1}  V(u/2)}{\partial ^{2r+1}u} \,T_M(u,v).
\end{equation}
Finally, we recognize that the infinite sum along $r$ can be cast into the difference between two shifted potentials, that is,
\begin{equation}
V\left(\frac{u+v}{2}\right)-V\left(\frac{u-v}{2}\right)=\sum_{r=0}^{\infty}\frac{v^{2r+1}}{2^{2r}(2r+1)!}  \frac{\partial^{2r+1}  V(u/2)}{\partial ^{2r+1}u}.
\end{equation}

Thus, we finally arrive at the following partial differential equation
\begin{equation}
\frac{2\hbar^2}{\mu}\frac{\partial^2 T_M(u,v)}{\partial u\partial v}=\,\left(V\left(\frac{u+v}{2}\right)-V\left(\frac{u-v}{2}\right)\right)  T_M(u,v),
\end{equation}
which is exactly the time kernel equation in $(u,v)$ coordinates. Going back to the original $(q,q')$ coordinates, we rederive the original TKE given by Eq. (\ref{TKE2}) with the corresponding boundary conditions $T_M(q,q)=q/2$ and $T_M(q,-q)=0.$ Since the TKE admits only one unique solution \cite{Sombillo2012,Farrales2022}, we then conclude $T_M(q,q')=T_S(q,q')$ so that $\hat{\mathrm{T}}_M = \hat{\mathrm{T}}_S$. This proves the isomorphism between the quantum phase space approach by Weyl, Wigner, and Moyal and the theory of supraquantization within the quantum TOA problem. 

\section{Examples}\label{sec:examples}

Let us now explore specific arrival time problems with two goals in mind. Firstly, we demonstrate how physical insights can be extracted from our Moyal TOA using phase space techniques, focusing on the time of arrival of a free particle. Secondly, we illustrate the construction of the Moyal TOA for nonlinear systems using a quartic anharmonic oscillator potential and highlight the quantum corrections to the corresponding classical arrival time.

\subsection{The time of arrival problem for a free-particle}
Let us revisit the time of arrival of a free particle. We assume a quantum particle initially located at $(q,p)$ in phase space with energy $H_F=p^2/2\mu$. We ask for the average value of measured arrival times at the origin.

We start by determining the associated TOA phase space function. For $V(q)=0$, the Moyal TOA readily simplifies to $\mathcal{T}_F(q,p)=-\mu q/p.$ It is straightforward to show that  $\mathcal{T}_F(q,p)$ satisfies the canonical relations $\{H_F(q,p),\mathcal{T}_F(q,p)\big\}_{M}=\{H_F(q,p),\mathcal{T}_F(q,p)\big\}_{PB}=1.$ 

In phase space formalism, the expectation value of an observable $\mathcal{A}$ appears simply as the integral of the phase space function $\mathcal{A}(q,p)$ over all $q$ and $p$ spaces weighted by the Wigner function $W_\psi(q,p)$ characterizing the particle's initial state \cite{Zachos2005}. Hence, given the function $\mathcal{T}_F(q,p)$, the corresponding expectation value $\left\langle \mathcal{T}_F \right\rangle$ appears as
\begin{equation}\label{freexpec}
\left\langle \mathcal{T}_F \right\rangle=\int_{-\infty}^{\infty} dp\int_{-\infty}^{\infty} dq\,\mathcal{T}_F(q,p)\,W(q,p),
\end{equation}
where the associated Wigner function is defined by
\begin{equation}\label{wignerq}
\begin{split}
W_\psi(q,p)&=\frac{1}{\pi \hbar} \int_{-\infty}^{\infty}dy \, \psi^*(q+y)\, \psi(q-y) \, e^{2ipy/\hbar},
\end{split}
\end{equation}
for an initial state $\psi(q)$. 

Let us now consider an incident particle represented by a Gaussian wave packet of the form 
\begin{equation}
\psi(q)=\frac{1}{\sqrt{\sigma\sqrt{2\pi}}}e^{-(q-q_o)^2/4\sigma^2}e^{ik_oq},
\end{equation}
with a momentum expectation value $\hbar k_0$, group velocity $\hbar k_0/\mu$, initial position $q_0$, momentum $p_0$, and position variance $\sigma^2$. The corresponding Wigner function is 
obtained by direct substitution of $\psi(q)$ into Eq. (\ref{wignerq}) leading to
\begin{equation}\label{wignergauss}
W_\psi(q,p)=\frac{1}{\pi \hbar}\,e^{-\frac{(q-q_o)^2}{2\sigma^2}}\,e^{-\frac{2\sigma^2}{\hbar^2}(p-\hbar k_o)^2}.
\end{equation}
Notice that the Wigner function is also of a Gaussian form and is positive definite. Insertion of $W_\psi(q,p)$ into Eq. (\ref{freexpec}), we find the following free TOA expectation value
\begin{equation}\label{freexpec2}
\left\langle \mathcal{T}_F\right\rangle=-\frac{\mu}{\hbar}\int_{-\infty}^{\infty} dp\,\frac{e^{-\frac{2\sigma^2}{\hbar^2}(p-\hbar k_o)^2}}{p}\int_{-\infty}^{\infty} dq\,q\,e^{-\frac{(q-q_o)^2}{2\sigma^2}}.
\end{equation}

The integral along $q$ is just a Gaussian integral and can be evaluated immediately. On the other hand, the divergent integral along $p$ is treated as a distribution by taking its principal value. Upon using the integral representation of the imaginary error function, $\mathrm{erfi}(x)=-(e^{x^2}/\pi)\,\mathrm{PV}\,\int_{-\infty}^{\infty}ds\,e^{-s^2}(s-x)^{-1},$ and its parity property, $\mathrm{erfi}(-x)=-\mathrm{erfi}(x)$, we finally obtain
\begin{equation}\label{expectTOA}
\left\langle \mathcal{T}_F \right\rangle=-\frac{\mu q_o}{p_o} \sqrt{2\pi}\,k_o \, \sigma \, e^{-2\sigma^2\,k_o^2}\,\,\mathrm{erfi}(\sqrt{2}\sigma k_o).
\end{equation}
Equation (\ref{expectTOA}) gives the expected quantum arrival time of a free particle at the origin. Notice that we can identify the factor $\mathcal{T}_{cl}(q_0,p_0)=-\mu\,q_0/p_0$ as the free classical TOA with initial position $q_0$ and momentum $p_0$, while the factor $Q(k_0\sigma)=\sqrt{2\pi}\,k_o \sigma \, e^{-2\,k_o^2\sigma^2}\,\,\mathrm{erfi}(\sqrt{2}\,k_o \sigma )$ as quantum correction. This correction depends only on \(k_0\) and \(\sigma\), describing the initial preparation of the incident particle. Hence, Eq. (\ref{expectTOA}) simplifies to $\left\langle \mathcal{T}_F\right\rangle=\mathcal{T}_{cl}(q_0,p_0)\,Q(k_0\sigma). $ This result exactly matches the findings of Ref. \cite{Galapon2009} using the theory of TOA operators. 

Our findings have also an intriguing implication regarding the nature of the Wigner function $W_\psi(q,p)$. It has been argued before that a positive everywhere Wigner function can only exhibit classical phenomena \cite{case2008}. However, this is not always true. In the context of quantum TOA, we have shown that a positive-definite Gaussian Wigner function describes quantum behavior for a free particle, leading to quantum corrections to the classical arrival time.

We highlight the importance of our phase space approach, which not only allows for comparison but also provides an avenue to analytically verify results obtained from the theory of quantized and supraquantized TOA operators. In fact, we arrive at Eq. (\ref{expectTOA}) much easier than in Ref. \cite{Galapon2009} since all we need to do is to perform a phase space integral with respect to the phase space function $\mathcal{T}_F(q,p)$ and Wigner function $W_\psi(q,p)$. In general, the method described in this subsection can be applied to any interaction potential. One starts by solving the Moyal TOA $\mathcal{T}_M(q,p)$ for a given potential $V(q)$. The average value of measured arrival times can be determined from the phase space integral defined in Eq. (\ref{freexpec}) for a given Wigner function $W_\psi(q,p).$

\subsection{The time of arrival problem for a quartic oscillator potential}

Let us also consider a quartic oscillator potential of the form $V(q)=\lambda q^4$ for some constant $\lambda \in \mathbb{R}$. This potential clearly yields a nonlinear equation of motion so that the quantum corrections $\hbar^{2n} \tau_{M,n}(q, p)$ discussed in Section (\ref{sec:mbtoa}) do not vanish. We wish to solve for these quantum corrections in phase space.

Following Eq. (\ref{toaexpandph}), the Moyal TOA appears as
\begin{equation}\label{toaexpandphquartic}
\mathcal{T}_M(q,p)=\tau_{M,0}(q,p)+\sum_{n=1}^{\infty}\hbar^{2n}\,\tau_{M,n}(q,p),
\end{equation}
where the leading term is the local TOA (Eq. (\ref{ltoapb}))
\begin{equation}\label{ltoapbquart}
\tau_{M,0}(q,p)=-\sum_{k=0}^{\infty}(-1)^k \frac{(2k-1)!!}{k!} \frac{\mu^{k+1}}{p^{2k+1}}\int_{0}^{q}dq' \, \left(V(q)-V(q')\right)^k.
\end{equation}
On the other hand, since the factor $ \partial^{2r+1} \, V(q')/\partial q'^{2r+1}$ in Eq. (\ref{tnqpfinal}) vanishes for all $r \ge 2$ for the quartic potential, the quantum corrections $\hbar^{2n}\tau_{M,n}(q,p)$ simplifies to
\begin{equation}\label{tnqpfinalquart}
\begin{split}
\hbar^{2n}\tau_{M,n}(q,p)&=-\frac{\mu}{24}\,\hbar^{2n}\int_{0}^{q}dq'\,\mathrm{exp}\left[\big(V(q)-V(q')\big)\frac{\mu}{p}\frac{\partial}{\partial p}\right]\frac{1}{p} \frac{\partial^{3} \, V(q')}{\partial q'^{3}}\, \frac{\partial^{3} \, \tau_{M,n-1}(q',p)}{\partial p^{3}}.
\end{split}
\end{equation}
The action of $\mathrm{exp}\left[\big(V(q)-V(q')\big)(\mu/p)(\partial/\partial p\right)]$ is determined by expanding the exponential function into an infinite series and then performing the partial derivatives on the subsequent factor with respect to $p$.

\subsection{The classical time of arrival}

Direct substitution of $V(q)=\lambda q^4$ into Eq. \eqref{ltoapbquart} and performing the resulting integral along $q'$ gives the following local TOA
\begin{equation}\label{ltoaquartic}
\tau_{M,0}(q,p)=-\frac{\mu q}{p} \,\sum_{k=0}^{\infty} \frac{(2k)!(-1)^k}{(5/4)_k \, k!}\left(\frac{\mu \lambda q^4}{2p^2}\right)^k. 
\end{equation}
When $2\mu\lambda q^4/p^2 <1$, the above equation converges absolutely to the classical time of arrival at the origin given by
\begin{equation}\label{tau0quartic}
\mathcal{T}_C(q,p)=-\frac{\mu q}{p}\,\,{}_2F_1\left(\frac{1}{2},1;\frac{5}{4};-\frac{2\mu\lambda q^4}{p^2}\right).
\end{equation}
Note that the hypergeometric function ${}_2F_1\left(a,b;c;z\right)$ has a branch point at $z=1$ and becomes complex-valued for $z>1$. This suggests that the factor ${}_2F_1\left(1/2,1;5/4;-2\mu\lambda q^4/p^2\right)$ is well-defined in the entire phase space when $\lambda>1$, ensuring that the corresponding classical arrival time $\mathcal{T}_C(q,p)$ is always real-valued. 

For $\lambda<1$, the argument $-2\mu\lambda q^4/p^2$ is always positive, so that ${}_2F_1\left(1/2,1;5/4;-2\mu\lambda q^4/p^2\right)$ can now become complex-valued. Consequently, the corresponding classical arrival time $\mathcal{T}_C(q,p)$ is no longer defined in the entire phase space. For this case, $\mathcal{T}_C(q,p)$ is only real-valued when $2\mu|\lambda | q^4/p^2<1$, otherwise, it is complex and the associated LTOA $\tau_{M,0}(q,p)$ diverges. For example, let $E=p^2/2\mu-|\lambda|q^4$ denote the initial total energy of the classical particle for the case $\lambda <1$. When $2\mu|\lambda | q^4/p^2<1$, the interaction potential decreases the kinetic energy of the particle, but the total energy remains positive, allowing the particle to reach the origin. Hence, we obtain a measurable time of arrival. Conversely, when $2\mu|\lambda | q^4/p^2>1$, the magnitude of the interaction potential exceeds the particle's kinetic energy, resulting in a negative total initial energy. This indicates that the particle is already bounded by the interaction potential from the beginning. As a result, the classical arrival time $\mathcal{T}_C(q,p)$ becomes complex, and the local TOA $\tau_{M,0}(q,p)$ becomes infinite, both indicating non-arrival at the origin. Hence, when $\lambda <1$, the further the initial position of the particle, the more negative its energy, and the tighter it is bounded by the potential. In such cases, no arrival time is measured.

\subsection{The first three quantum corrections}

Let us also calculate the first three quantum corrections to the classical arrival time dependent on $\hbar^2$,  $\hbar^4$, and  $\hbar^6$. 

Setting $n=1$ in Eq. (\ref{tnqpfinalquart}), expanding the exponential function in series form, and performing the indicated partial derivatives and integral along $q'$, one arrives at the following expansion
\begin{equation}\label{2ndcorrecquart}
\hbar^{2}\,\tau_{M,1}(q,p)=-\mu^2 \lambda\,\frac{q^3}{p^5}\,\hbar^2\,\sum_{k=0}^\infty \frac{(-1)^k \,\Gamma \left(k+5/2\right)}{\sqrt{\pi}\,\Gamma\left(k+7/4\right)}\left(\frac{2\mu\lambda q^4}{p^2}\right)^k \sum_{l=0}^k \frac{(2l+2)\Gamma(l+3/4)}{(5/4)_l},
\end{equation}
The function $\hbar^{2}\,\tau_{M,1}(q,p)$ serves as the leading quantum correction to the local TOA \eqref{ltoaquartic}.  It also converges when $2\mu\lambda q^4/p^2 <1$. When the convergence condition is met, $\hbar^{2}\,\tau_{M,1}(q,p)$ converges absolutely to the phase space function $\mathcal{T}_1(q,p,\hbar^2)$ given by
\begin{equation}\label{t1qpquartic}
\mathcal{T}_1(q,p,\hbar^2)=-\mu^2 \lambda\,\frac{q^3}{p^5}\,\hbar^2\left[\frac{5}{2}\,\,{}_2F_1\left(1,\frac{7}{2};\frac{5}{4};-\frac{2\mu\lambda q^4}{p^2}\right)-\frac{1}{2}\,\,{}_2F_1\left(1,\frac{5}{2};\frac{7}{4};-\frac{2\mu\lambda q^4}{p^2}\right)\right],
\end{equation}
with the inclusion $\hbar^2\tau_{M,1}(q,p) \subset \mathcal{T}_{1}(q,p,\hbar^2)$. In other words, $\mathcal{T}_1(q,p,\hbar^2)$ is the global form of $\hbar^{2}\,\tau_{M,1}(q,p)$ within the latter's region of convergence in phase space. Hence, the phase space function $\mathcal{T}_1(q,p,\hbar^2)$ serves as the leading quantum correction factor to the classical time of arrival \eqref{tau0quartic}. See also Appendix (\ref{app:quarticcorrections}) for the derivation of Eqs. (\ref{2ndcorrecquart}) and (\ref{t1qpquartic}).

Performing similar calculations for $n=2$ and $n=3$ in Eq. (\ref{tnqpfinalquart}), we arrive at the global form of the next two quantum corrections to the classical arrival time $\mathcal{T}_C(q,p)$ given by
\begin{equation}
\begin{split}
\mathcal{T}_2(q,p,\hbar^4)=-\mu^3 \lambda^2\frac{q^5}{p^9}\hbar^4&\bigg[\frac{14}{3}\,{}_4F_3\left(2,2,2,\frac{9}{2}\,;1,1,\frac{9}{4}; -\frac{2\mu \lambda\,q^4}{p^2}\right)+\frac{301}{3}\, {}_3F_2\left(2,\frac{113}{27},\frac{9}{2}\,;\frac{9}{4},\frac{86}{27}; -\frac{2\mu \lambda\,q^4}{p^2}\right)\\
&-\frac{175}{4}\,\, {}_3F_2\left(1,\frac{9}{2},\frac{17}{2}\,;\frac{7}{4},\frac{15}{2}\,; -\frac{2\mu \lambda\,q^4}{p^2}\right)+\frac{91}{4}\,\, {}_2F_1\left(1,\frac{9}{2}\,\,;\,\frac{9}{4}\,;-\frac{2\mu \lambda\,q^4}{p^2}\right)\bigg],
\end{split}
\end{equation}
\begin{equation}
\begin{split}
\mathcal{T}_3(q,p,&\hbar^6)=-\mu^4 \lambda^3\frac{q^7}{p^{13}}\hbar^6\bigg[\frac{1166}{3}{}_5F_4\left(2,2,2,\frac{60}{7},\frac{13}{2};1,1,\frac{9}{4},\frac{53}{7};-\frac{2\mu \lambda q^4}{p^2}\right)\\
&-\frac{154}{5}{}_2F_1\left(1,\frac{13}{2};\frac{9}{4};-\frac{2\mu\lambda q^4}{p^2}\right)+\frac{891}{2}\, {}_4F_3\left(2,2,2,\frac{13}{2};1,1,\frac{9}{4};-\frac{2\mu \lambda q^4}{p^2}\right)\\
&-55\, {}_4F_3\left(2,2,2,\frac{13}{2};1,1,\frac{11}{4};-\frac{2\mu \lambda q^4}{p^2}\right)+\frac{54131}{12}\, {}_3F_2\left(2,2,\frac{13}{2};1,\frac{9}{4};-\frac{2\mu \lambda q^4}{p^2}\right)\\
&+\frac{284889}{40}\, {}_3F_2\left(1,\frac{21277}{12644},\frac{13}{2};\frac{8633}{12644},\frac{9}{4};-\frac{2\mu \lambda q^4}{p^2}\right)-\frac{12265}{2}\, {}_3F_2\left(2,\frac{515}{69},\frac{13}{2};\frac{11}{4},\frac{446}{69};-\frac{2\mu \lambda q^4}{p^2}\right)\\
&+\frac{75075}{8}\, {}_3F_2\left(1,\frac{13}{2},\frac{27}{2};\frac{9}{4},\frac{25}{2};-\frac{2\mu \lambda q^4}{p^2}\right)-\frac{15125}{4}\,{}_2F_1\left(1,\frac{13}{2};\frac{11}{4};-\frac{2\mu \lambda q^4}{p^2}\right)\\
&+\frac{418}{15}\,{}_2F_1\left(2,\frac{13}{2};\frac{9}{4};-\frac{2\mu \lambda q^4}{p^2}\right)\bigg].
\end{split}
\end{equation}
The expansion of $\mathcal{T}_2(q,p,\hbar^4)$ and $\mathcal{T}_3(q,p,\hbar^6)$ around the free TOA at the origin constitute the second and third quantum corrections, $\hbar^{4}\,\tau_{M,2}(q,p)$ and $\hbar^{6}\,\tau_{M,3}(q,p)$, to the local time of arrival $\tau_{M,0}(q,p)$ \eqref{ltoapbquart}, respectively. In the classical limit $\hbar \to 0$, their contributions identically vanish so that the Moyal time of arrival correctly leads to the correct classical arrival time.

We note that all quantum correction terms $\mathcal{T}_n(q,p,\hbar^{2n})$ for the quartic oscillator potential can be written as sums of hypergeometric functions of the form ${}_{p+1}F_p(a_1,...,a_{p+1};b_1,...,b_p;z)$. This observation follows from Eq. (\ref{tnqpfinalquart}), which involves specific integrals and derivatives of the previous iterate, also of the form ${}_{p+1}F_p(a_1,...,a_{p+1};b_1,...,b_p;z)$. Upon using the known differentiation of the hypergeometric function  \cite{derivativepfq}
\begin{equation}
\frac{\partial^n\left(z^\alpha \,{}_pF_q(a_1,...,a_{p};b_1,...,b_p;z)\right)}{\partial z^n}=(-1)^n (-\alpha)_n\,z^{\alpha-n}\,{}_{p+1}F_{q+1}(\alpha+1,a_1,...,a_{p};\alpha-n+1,b_1,...,b_p;z),
\end{equation}
for $n \in \mathbb{N}^+$, and the following indefinite integral \cite{intpfq}
\begin{equation}
 \int dz \, z^{\alpha-1}   \, \,{}_pF_q(a_1,...,a_{p};b_1,...,b_p;z) = \frac{z^\alpha}{\alpha}\,{}_{p+1}F_{q+1}(\alpha,a_1,...,a_{p};\alpha+1,b_1,...,b_p;z),
\end{equation}
and noting that the initial condition $\mathcal{T}_C(q,p)$ is written in hypergeometric function ${}_2F_1\left(a_1,a_2;b_1;z\right)$, we can establish that all the succeeding quantum corrections are of the form ${}_{p+1}F_p(a_1,...,a_{p+1};b_1,...,b_p;z)$.

Hence, the analysis done for $\mathcal{T}_C(q,p)$ applies similarly to the quantum correction terms $\mathcal{T}_n(q,p,\hbar^{2n})$, in particular, to the first three quantum corrections described above. They all share branch points at $z=1$ and become complex-valued when $z>1$, similar to ${}_2F_1\left(a,b;c;z\right)$. When $\lambda >1$, these quantum corrections are well-defined across the entire phase space. However, when $\lambda <1$ and $2\mu|\lambda | q^4/p^2>1$, the quantum corrections become complex-valued, and their local expansions diverge, similar to $\mathcal{T}_C(q,p)$ and $\tau_{M,0}(q,p)$, respectively.

Lastly, the approximated Moyal TOA up to the third quantum correction, $\mathcal{T}_M(q,p)\approx\tau_{M,0}(q,p)+\hbar^2\tau_{M,1}(q,p)+\hbar^4\tau_{M,2}(q,p)+\hbar^6\tau_{M,3}(q,p)$, exactly correspond to the inverse Weyl-Wigner transform of the time kernel of the approximated supraquantized TOA operator, $\hat{\mathrm{T}}_S \approx \hat{\mathrm{T}}_W+ \hat{\mathrm{T}}_1+ \hat{\mathrm{T}}_2+\hat{\mathrm{T}}_3 $, for the quartic oscillator potential discussed in Ref. \cite{Pablico2023}. This demonstrates the correspondence between the Moyal TOA and the supraquantized TOA operator for this specific potential in accordance to Eq. \eqref{inversefor}.

\section{Conclusions}\label{sec:conclusions}
We have considered the quantum arrival time problem of a structureless particle within the phase space formulation of quantum theory by Weyl, Wigner, and Moyal with three objectives in mind. 

Our first objective was to introduce an alternative framework for obtaining quantum images of the classical arrival time independent of quantization and operator-based formulations. This is facilitated by rewriting the canonical commutation relation $[\hat{\mathrm{H}},\hat{\mathrm{T}}]=i\hbar \mathbb{1}$ as the Moyal bracket between time and energy, that is, $\big\{H(q,p),\mathcal{T}_M(q,p)\big\}_{MB}=1.$ Imposing the necessary conditions of arrival time observables, we then showed that the required quantum image $\mathcal{T}_M(q,p)$ is a real-valued and time-reversal symmetric function in formal series in the deformation parameter $\hbar^2$, that is, $\mathcal{T}_M(q,p)=\tau_{M,0}(q,p)+\sum_{n=1}^{\infty}\hbar^{2n}\,\tau_{M,n}(q,p)$. The leading term coincides with the classical arrival time and the succeeding terms can be obtained recursively in closed-form from Eq. (\ref{tnqpfinal}). The $\hbar$-dependent terms vanish for linear systems but are generally non-vanishing for nonlinear systems. Since the Moyal bracket is a formal deformation of the Poisson bracket, we have also interpreted $\mathcal{T}_M(q,p)$ as the Moyal deformation of the classical arrival time. Our method could be extended to obtaining quantum images of other observables satisfying different commutation relation. 

Our second objective was to test the theoretical predictions of the theory of supraquantization introduced in Refs. \cite{Galapon2004,Pablico2023} by comparison with the current phase space approach. This was done by taking the Weyl map of $\mathcal{T}_M(q,p)$ to determine the corresponding operator $\hat{\mathrm{T}}_M$, which we have called the Moyal TOA operator. Within the rigged Hilbert space formulation of quantum theory, the Moyal operator $\hat{\mathrm{T}}_M$ is exactly equal to the supraquantized TOA operator $\hat{\mathrm{T}}_S$ constructed independently of canonical quantization. The equality of the two operators highlights the isomorphism between the quantum phase space and the theory of supraquantization. 

Finally, our third objective was to identify possible modifications to the theory of quantized TOA operators if one wishes to incorporate the correct algebra of time observables. Our results suggest that the time-energy conjugacy relation should be represented by the Moyal bracket, not the Poisson bracket, in phase space. Moreover, the equality (inequality) of the Weyl-quantized and supraquantized TOA operators for linear (nonlinear) systems is due to the equality (inequality) of the corresponding Poisson and Moyal brackets. Hence, if we wish to incorporate the proper algebra of arrival time observables, the Weyl quantization of the classical arrival time $\mathcal{T}_C(q,p)$ must be replaced by the Weyl mapping of the Moyal arrival time $\mathcal{T}_M(q,p)$. 

We then analyzed two specific arrival time problems involving a free particle and a quartic oscillator potential, where we have demonstrated how to construct and extract the physical significance of the Moyal TOA and the subsequent quantum corrections to the classical arrival time.

Currently, the direct implementation of the supraquantized time-of-arrival (TOA) operator, as well as the Moyal TOA phase space function, in experimental measurements of time remains an open problem. Nevertheless, the supraquantized operator has been successfully applied to the quantum tunneling time problem, predicting zero tunneling time \cite{galapsbarrier,arxivflores}, a result that is consistent with various experimental studies \cite{ecklehelium,sainadh}. While this provides some support for the operator's validity, and consequently for the Moyal TOA, further research is still needed to fully explore the experimental applicability of these quantum images of the classical TOA.

\section{Acknowledgment}\label{sec:acknow}
D.A.L. Pablico gratefully acknowledges the support of the Department of Science and Technology - Science Education Institute (DOST-SEI) through the Accelerated Science and Technology Human Resource Development Program (ASTHRDP) graduate scholarship program.

\appendix
\section*{Appendix}
\renewcommand\thesubsection{\Alph{subsection}}
\renewcommand{\theequation}{\Alph{subsection}.\arabic{equation}}
\setcounter{subsection}{0}

\subsection{Derivation and validation of the LTOA from the Poisson bracket relation with the Hamiltonian}\label{appendix1}

\setcounter{equation}{0}

We start from the recurrence relation given by Eq. (\ref{succtqpn}), 
\begin{equation}\label{app:rec}
\mathcal{T}_{C,n}(q,p)=-\frac{\mu q}{p}+\frac{\mu}{p}\int_{0}^{q} dq'\frac{\partial V(q')}{\partial q'}\frac{\partial \mathcal{T}_{C,n-1}(q',p)}{\partial p}.
\end{equation}
For $n=1$, we find
\begin{equation}
\mathcal{T}_{C,1}(q,p)=-\frac{\mu q}{p}+\frac{\mu^2}{p^3}\int_{0}^{q} dq'\,q'\,\frac{\partial V(q')}{\partial q'},
\end{equation}
where we have used the initial condition $\mathcal{T}_{C,0}(q,p)=-\mu q/p$. Integration by parts leads to
\begin{equation}\label{app:tc1}
\mathcal{T}_{C,1}(q,p)=-\frac{\mu q}{p}+\frac{\mu^2}{p^3}\int_{0}^{q} dq'\,\left(V(q)-V(q')\right).
\end{equation}

For $n=2$, Eqs. (\ref{app:rec}) and (\ref{app:tc1}) give
\begin{equation}
\mathcal{T}_{C,2}(q,p)=-\frac{\mu q}{p}+\frac{\mu^2}{p^3}\int_{0}^{q} dq'\,\left(V(q)-V(q')\right)-\frac{3\mu^3}{p^5}\int_{0}^{q} dq'\,\frac{\partial V(q')}{\partial q'}\int_{0}^{q'} dq''\,\left(V(q')-V(q'')\right).
\end{equation}
We interchange the order of integrations using the following identity
\begin{equation}\label{app:ident}
\int_{0}^{\alpha}dx\int_{0}^{x}dy\, f(x,y)=\int_{0}^{\alpha}dy\int_{y}^{\alpha}dx\, f(x,y),
\end{equation}
and rewrite the integrand using the relation
\begin{equation}\label{app:deriV}
\frac{\partial}{\partial q'}\big[V(q')-V(q'')\big]^n =n\big[V(q')-V(q'')\big]^{n-1}\,\frac{\partial V(q')}{\partial q'}.
\end{equation}

The double integral could be simplified to get 
\begin{equation}\label{app:tc2}
\mathcal{T}_{C,2}(q,p)=-\frac{\mu q}{p}+\frac{\mu^2}{p^3}\int_{0}^{q} dq'\,\left(V(q)-V(q')\right)-\frac{3\mu^3}{p^5}\int_{0}^{q} dq'\,\left(V(q)-V(q')\right)^2.
\end{equation}

Note that the interchange is valid due to the continuity of the potential $V(q)$ and the convergence of the integrals. A similar argument is given every time we interchange the order integrations in the current section. 

For $n=3$, substitution of Eq. (\ref{app:tc2}) into (\ref{app:rec}) gives
\begin{align}\label{app:tc3}
\mathcal{T}_{C,3}(q,p)=&-\frac{\mu q}{p}+\frac{\mu^2}{p^3}\int_{0}^{q} dq'\,\left(V(q)-V(q')\right)-\frac{3\mu^3}{p^5}\int_{0}^{q} dq'\,\left(V(q)-V(q')\right)^2\\
&+\frac{15\mu^4}{2p^7}\int_{0}^{q} dq'\,\frac{\partial V(q')}{\partial q'}\int_{0}^{q'} dq''\,\left(V(q')-V(q'')\right)^2.
\end{align}
Again, we interchange the order of integrations in accordance to Eq. (\ref{app:ident}), rewrite the integrand using Eq. (\ref{app:deriV}), then perform the indicated operations, we find
\begin{align}\label{app:tc3fin}
\mathcal{T}_{C,3}(q,p)=&-\frac{\mu q}{p}+\frac{\mu^2}{p^3}\int_{0}^{q} dq'\,\left(V(q)-V(q')\right)-\frac{3\mu^3}{p^5}\int_{0}^{q} dq'\,\left(V(q)-V(q')\right)^2\\
&+\frac{5\mu^4}{2p^7}\int_{0}^{q} dq'\,\frac{\partial V(q')}{\partial q'}\int_{0}^{q} dq'\,\left(V(q)-V(q')\right)^3.
\end{align}

Repeating similar steps for higher $n$'s, we identify the following general form
\begin{equation}\label{app:tcn}
\mathcal{T}_{C,n}(q,p)=-\sum_{k=0}^{n}(-1)^k \frac{(2k-1)!!}{k!} \frac{\mu^{k+1}}{p^{2k+1}}\int_{0}^{q}dq' \, \left(V(q)-V(q')\right)^k,
\end{equation}
for $n \ge 1$. Taking the limit $n\to\infty$ immediately leads to the LTOA given by Eq. (\ref{ltoapb}). 

For completeness, we validate Eq. (\ref{app:tcn}) by mathematical induction. We assume that the above equation is valid for some $n=j \ge 1$ and show that it still holds true for the next iterate $n=j+1$. For $n=j+1$, Eq. (\ref{app:tcn}) becomes
\begin{equation}\label{apptcj1}
\mathcal{T}_{C,j+1}(q,p)=-\sum_{k=0}^{j+1}(-1)^k \frac{(2k-1)!!}{k!} \frac{\mu^{k+1}}{p^{2k+1}}\int_{0}^{q}dq' \, \left(V(q)-V(q')\right)^k. 
\end{equation}
Isolating the leading term, $k=0$, and shifting index from $k$ to $k-1$ leads to
\begin{equation}\label{app:tcj+111}
\mathcal{T}_{C,j+1}(q,p)=-\frac{\mu q}{p}+\frac{\mu}{p}\sum_{k=0}^{j}(-1)^k \frac{(2k+1)!!}{(k+1)!} \frac{\mu^{k+1}}{p^{2k+2}}\int_{0}^{q}dq' \, \left(V(q)-V(q')\right)^{k+1}.
\end{equation}
Our goal is to obtain the correct recurrence relation for $\mathcal{T}_{C,j+1}(q,p)$ in accordance to Eq. (\ref{app:rec}). Notice that the integrand can be rewritten as some specific integral given by
\begin{equation}\label{app:idenint}
\left(V(q)-V(q')\right)^{k+1}=\int_{q'}^{q}dq'' \, \frac{\partial }{\partial q''}\left[\left(V(q'')-V(q')\right)^{k+1} \right]. 
\end{equation}

Using Eq. (\ref{app:ident}) and (\ref{app:idenint}), followed by some straightforward simplifications, Eq. (\ref{app:tcj+111}) can be rewritten as 
\begin{equation}\label{app:tcj+12}
\mathcal{T}_{C,j+1}(q,p)=-\frac{\mu q}{p}+\frac{\mu}{p}\int_{0}^{q}dq''\,\frac{\partial V(q'')}{\partial q''} \,\frac{\partial}{\partial p}\left[- \,\sum_{k=0}^{j}(-1)^k \frac{(2k-1)!!}{k!} \frac{\mu^{k+1}}{p^{2k+1}}\int_{0}^{q''}dq'\, \left(V(q'')-V(q')\right)^{k}\right].
\end{equation}
By comparison with Eq. (\ref{app:tcn}), the factor inside the square brackets is identified as $\mathcal{T}_{C,j}(q'',p)$. Hence, we finally find,
\begin{equation}\label{app:recfin}
\mathcal{T}_{C,j+1}(q,p)=-\frac{\mu q}{p}+\frac{\mu}{p}\int_{0}^{q} dq''\frac{\partial V(q'')}{\partial q''}\frac{\partial \mathcal{T}_{C,j}(q'',p)}{\partial p},
\end{equation}
which is the correct recurrence relation for the $n=j+1$ case, in agreement with Eq. (\ref{app:rec}).

\subsection{Derivation and validation of the Moyal deformation of the LTOA }\label{appendix2}
\setcounter{equation}{0}

Recall Eq. (\ref{iteratetnm}) given by
\begin{equation}\label{app:iteratetnm}
F_{m}(q,p)=F_{0}(q,p)+\frac{\mu}{p}\int_{0}^{q}dq' \, \frac{\partial V(q')}{\partial q'}\,\frac{\partial F_{m-1}(q',p)}{\partial p},
\end{equation}
with the initial condition
\begin{equation}\label{app:iteratetIC}
F_{0}(q,p)=\frac{\mu}{p}\,\sum_{r=1}^{2n}\,\frac{(-1)^r}{2^{2r}\,(2r+1)!} \,\int_{0}^{q}dq'\, \frac{\partial^{2r+1} \, V(q')}{\partial q'^{2r+1}}\, \frac{\partial^{2r+1} \, \tau_{M,n-r}(q',p) }{\partial p^{2r+1}}.
\end{equation}

We want to solve Eq. (\ref{app:iteratetnm}). To start, we set $m=1$ in Eq. (\ref{app:iteratetnm}) followed by the substitution of $F_{0}(q,p)$. The resulting double integral can be simplified with the help of the identities defined by Eqs. (\ref{app:ident}) and (\ref{app:deriV}). We get the following first-order approximation
\begin{equation}\label{app2:taun1}
F_{1}(q,p)=F_{0}(q,p)+\frac{\mu}{p}\frac{\partial}{\partial p}\left[\frac{\mu}{p}\sum_{r=1}^{2n}\frac{(-1)^r}{2^{2r}(2r+1)!} \,\int_{0}^{q}dq'\big(V(q)-V(q')\big) \frac{\partial^{2r+1} \, V(q')}{\partial q'^{2r+1}}\frac{\partial^{2r+1} \, \tau_{M,n-r}(q',p) }{\partial p^{2r+1}}\right].
\end{equation}
This equation can be further simplified to
\begin{equation}\label{app2:taunf1}
F_{1}(q,p)= \sum_{j=0}^{1}\left(\frac{\mu}{p}\frac{\partial}{\partial p}\right)^j\frac{\mu}{p}\sum_{r=1}^{2n}\frac{(-1)^r}{2^{2r}(2r+1)!} \,\int_{0}^{q}dq'\big(V(q)-V(q')\big)^j\, \frac{\partial^{2r+1} \, V(q')}{\partial q'^{2r+1}}\frac{\partial^{2r+1} \, \tau_{M,n-r}(q',p) }{\partial p^{2r+1}}.
\end{equation}

For $m=2$, substitution of Eq. (\ref{app2:taun1}) into Eq. (\ref{app:iteratetnm}) gives 
\begin{equation}
\begin{split}
F_{2}(q,p)&=F_{1}(q,p)+\frac{\mu}{p}\frac{\partial}{\partial p}\bigg[\frac{\mu}{p}\frac{\partial}{\partial p}\bigg(\frac{\mu}{p}\sum_{r=1}^{2n}\frac{(-1)^r}{2^{2r}(2r+1)!} \,\frac{\partial^{2r+1}}{\partial p^{2r+1}}\int_{0}^{q}dq''\,\frac{\partial V(q'')}{\partial q''} \\ 
& \times\int_{0}^{q''}dq'\frac{\partial^{2r+1} \, V(q')}{\partial q'^{2r+1}}\,\big(V(q'')-V(q')\big)\,\tau_{M,n-r}(q',p)\bigg)\bigg].
\end{split}
\end{equation}
Again, interchanging the order of integrations and rewriting the integrand using Eqs. (\ref{app:ident}) and (\ref{app:deriV}), respectively, we find
\begin{equation}\label{app2:taun2}
F_{2}(q,p)= \sum_{j=0}^{2}\frac{1}{j!}\left(\frac{\mu}{p}\frac{\partial}{\partial p}\right)^j\frac{\mu}{p}\sum_{r=1}^{2n}\frac{(-1)^r}{2^{2r}(2r+1)!} \,\int_{0}^{q}dq'\big(V(q)-V(q')\big)^j\, \frac{\partial^{2r+1} \, V(q')}{\partial q'^{2r+1}}\frac{\partial^{2r+1} \, \tau_{M,n-r}(q',p) }{\partial p^{2r+1}}.
\end{equation}

Similar calculations for $m=3$ gives
\begin{equation}\label{app2:taun3}
F_{3}(q,p)= \sum_{j=0}^{3}\frac{1}{j!}\left(\frac{\mu}{p}\frac{\partial}{\partial p}\right)^j\frac{\mu}{p}\sum_{r=1}^{2n}\frac{(-1)^r}{2^{2r}(2r+1)!} \,\int_{0}^{q}dq'\big(V(q)-V(q')\big)^j\, \frac{\partial^{2r+1} \, V(q')}{\partial q'^{2r+1}}\frac{\partial^{2r+1} \, \tau_{M,n-r}(q',p) }{\partial p^{2r+1}}.
\end{equation}

A quick glance on Eqs. (\ref{app2:taun1}), (\ref{app2:taun2}), and (\ref{app2:taun3}) allow us infer the following general form
\begin{equation}\label{app2:taunm}
F_{m}(q,p)= \sum_{r=1}^{2n}\frac{(-1)^r}{2^{2r}(2r+1)!} \,\int_{0}^{q}dq'\sum_{j=0}^{m}\frac{1}{j!}\left(\frac{\mu}{p}\frac{\partial}{\partial p}\right)^j\frac{\mu}{p}\big(V(q)-V(q')\big)^j\, \frac{\partial^{2r+1} \, V(q')}{\partial q'^{2r+1}}\frac{\partial^{2r+1} \, \tau_{M,n-r}(q',p) }{\partial p^{2r+1}},
\end{equation}
which confirms Eq. (\ref{taunm}). The two finite sums are independent of each other so we can place the sum along $j$ inside the integral. In the limit $m\to \infty$, the series can be closed leading to Eq. (\ref{tnqpfinal}).  

For completeness, we also validate Eq. (\ref{app2:taunm}) similar to the methods of Appendix \ref{appendix1}. We assume that Eq. (\ref{app2:taunm}) is valid for $m=k$. We need to show that our expression is still valid for the case $m=k+1$. For the latter case, Eq. (\ref{app2:taunm}) gives

\begin{equation}\label{app2:taukm1}
F_{k+1}(q,p)= \sum_{r=1}^{2n}\frac{(-1)^r}{2^{2r}(2r+1)!} \int_{0}^{q}dq'\sum_{j=0}^{k+1}\frac{1}{j!}\left(\frac{\mu}{p}\frac{\partial}{\partial p}\right)^j\frac{\mu}{p}\big(V(q)-V(q')\big)^j\, \frac{\partial^{2r+1} \, V(q')}{\partial q'^{2r+1}}\frac{\partial^{2r+1} \tau_{M,n-r}(q',p) }{\partial p^{2r+1}}.
\end{equation}

We isolate the leading term, $j=0$, and shift index from $j$ to $j-1$. We find
\begin{equation}\label{app2:taukm1b}
\begin{split}
F_{k+1}(q,p)=F_0(q,p)\,+\,& \sum_{r=1}^{2n}\frac{(-1)^r}{2^{2r}(2r+1)!} \int_{0}^{q}dq'\sum_{j=0}^{k}\frac{1}{(j+1)!}\left(\frac{\mu}{p}\frac{\partial}{\partial p}\right)^{j+1}\,\frac{\mu}{p}\,\big(V(q)-V(q')\big)^{j+1}\\
&\times\frac{\partial^{2r+1} \, V(q')}{\partial q'^{2r+1}}\,\frac{\partial^{2r+1} \tau_{M,n-r}(q',p) }{\partial p^{2r+1}}.
\end{split}
\end{equation}

We use Eq. (\ref{app:idenint}) to rewrite the factor $\big(V(q)-V(q')\big)^{j+1}$ into an integral, followed by an interchange in the order of integrations in accordance to Eq. (\ref{app:ident}). After some rearrangements and evaluations, we derive
\begin{equation}
\begin{split}
F_{k+1}(q,p)=F_0(q,p)\,+\,&\int_{0}^{q}dq'\,\frac{\partial V(q'')}{\partial q''} \,\left(\frac{\mu}{p}\frac{\partial}{\partial p}\right) \bigg[ \sum_{r=1}^{2n}\frac{(-1)^r}{2^{2r}(2r+1)!} \int_{0}^{q''}dq'\sum_{j=0}^{k}\frac{1}{(j)!}\left(\frac{\mu}{p}\frac{\partial}{\partial p}\right)^{j}\,\frac{\mu}{p}\,\\
&\times\big(V(q)-V(q')\big)^{j}\frac{\partial^{2r+1} \, V(q')}{\partial q'^{2r+1}}\,\frac{\partial^{2r+1} \tau_{M,n-r}(q',p) }{\partial p^{2r+1}}\bigg].
\end{split}
\end{equation}

One immediately notice that the factor inside of the square brackets is simply $F_k(q,p)$ based on Eq. (\ref{app2:taunm}). The above equation then simplifies to
\begin{equation}\label{app2:iteratetnmk1}
F_{k+1}(q,p)=F_{0}(q,p)+\frac{\mu}{p}\int_{0}^{q}dq' \, \frac{\partial V(q')}{\partial q'}\,\frac{\partial F_{k}(q',p)}{\partial p},
\end{equation}
which is the correct recurrence relation for the $m=k+1$ case as indicated by Eq. (\ref{app:iteratetnm}).

\subsection{Validation of the time kernel factor corrections}\label{app:tkfn}

We validate the kernel $\langle q|\hat{\mathrm{T}}_{M,n}|q'\rangle =(\mu/i\hbar)\,\mathrm{sgn}(q-q')\,T_{M,n}(q,q')$ where
\begin{equation}
T_{M,n}(q,q')=\left(\frac{\mu}{2\hbar^2}\right)\sum_{r=1}^{n} \frac{1}{(2r+1)!}\frac{1}{2^{2r}}\int_{0}^{u} ds \, V^{(2r+1)}\left(\frac{s}{2}\right) \int_{0}^{v} dw \, w^{2r+1} \, T_{M,n-r}(s,w) \, G(s,w),
\end{equation}
with $u=q+q'$, $v=q-q'$, and
\begin{equation}
G(s,w)={}_0F_1 \left(;1;\left(\frac{\mu}{2\hbar^2}\right)(v^2-w^2)\left[V \left(\frac{u}{2}\right)-V \left(\frac{s}{2}\right)\right]\right).
\end{equation}

We assume $\langle q|\hat{\mathrm{T}}_{M,n}|q'\rangle$ to be valid for some $n=k$ and show that the resulting expression holds true for the next iterate $n=k+1$. For $n=k+1$, we have  $\langle q|\hat{\mathrm{T}}_{M,k+1}|q'\rangle=(\mu/i\hbar)\,\mathrm{sgn}(q-q')\,T_{M,k+1}(q,q')$ with the corresponding time kernel factor
\begin{equation}
T_{M,k+1}(q,q')=\left(\frac{\mu}{2\hbar^2}\right)\sum_{r=1}^{k+1} \frac{1}{(2r+1)!}\frac{1}{2^{2r}}\int_{0}^{u} ds \, V^{(2r+1)}\left(\frac{s}{2}\right) \int_{0}^{v} dw \, w^{2r+1} \, T_{M,k-r+1}(s,w) \, G(s,w).
\end{equation}

We take the inverse Weyl map of the above kernel using Eq. (\ref{oinverse}). This gives
\begin{equation}\label{tk1qpp}
	\begin{split}
\hbar^{2(k+1)}\tau_{M,k+1}\,(q,p)=\frac{\mu^2}{2i\hbar^3}\sum_{r=1}^{k+1} \frac{1}{(2r+1)!}\frac{1}{2^{2r}}\,&\int_{-\infty}^{\infty}dv\,\mathrm{sgn}(v)e^{-ivp/\hbar}\int_{0}^{2q}ds \frac{\partial^{2r+1} V\left(s/2\right)}{\partial q'^{2r+1}}\\
&\times\int_{0}^{v}dw\, w^{2r+1}\,T_{M,k+1-r}(s,w)\, G(s,w).
\end{split}
\end{equation}

The absolute convergence of $G(s,w)$ and the continuity of the interaction potential $V(q$) and its derivatives allow us to interchange safely the order of integrations along $s$ and $v$. We then perform integration by parts along the $v$-integral with $u=\mathrm{sgn}(v)\,F(v)$, where $F(v)$ is the whole integral along $w$, and $dv=e^{-ivp/\hbar}.$ Equation (\ref{tk1qpp}) evaluates to
\begin{equation}\label{tk1qpp2}
\begin{split}
\hbar^{2(k+1)}\tau_{M,k+1}\,(q,p)=-\frac{\mu^2}{2p\hbar^2}\sum_{r=0}^{k+1} \frac{1}{(2r+1)!}\frac{1}{2^{2r}}\,&\int_{0}^{2q}ds\, \frac{\partial^{2r+1} V\left(s/2\right)}{\partial q'^{2r+1}}\\
&\times\int_{-\infty}^{\infty}dv\,\mathrm{sgn}(v)e^{-ivp/\hbar}\,v^{2r+1}\,T_{M,k+1-r(s,v)}.
\end{split}
\end{equation}

By the Leibniz integral rule, we can rewrite the $v$-integral in terms of a derivative under an integral sign. Hence,
\begin{equation}\label{tk1qpp3}
\begin{split}
\hbar^{2(k+1)}\tau_{M,k+1}\,(q,p)=-\frac{\mu^2}{2p\hbar^2}\sum_{r=0}^{k+1} \frac{\left(i\hbar\right)^{2r+1}}{(2r+1)!}\frac{1}{2^{2r}}\,&\int_{0}^{2q}ds \frac{\partial^{2r+1} V\left(s/2\right)}{\partial q'^{2r+1}}\\
&\times\frac{\partial^{2r+1}}{\partial p^{2r+1}}\int_{-\infty}^{\infty}dv\,\mathrm{sgn}(v)e^{-ivp/\hbar}\,T_{M,k+1-r(s,v)}.
\end{split}
\end{equation}

We determine that the integral along $v$ can be rewritten in terms of the phase space function  $\hbar^{2(k+1-r)}\tau_{M,k+1-r}\,(q',p)$ following Eq. (\ref{oinverse}). Hence, we arrive at
\begin{equation}\label{app:tk1qpp4}
\tau_{M,k+1}\,(q,p)=\frac{\mu}{p}\sum_{r=0}^{k+1} \frac{\left(-1\right)^{r}}{(2r+1)!}\frac{1}{2^{2r}}\,\int_{0}^{2q}ds \frac{\partial^{2r+1} V\left(s/2\right)}{\partial q'^{2r+1}}\frac{\partial^{2r+1}\,\tau_{M,k+1-r}\,(q',p)}{\partial p^{2r+1}}.
\end{equation}
which is the known integral equation satisfied by $\tau_{M,k+1}\,(q,p)$ for $n=k+1$ in accordance to Eq. (\ref{wwtn}). Hence, the inferred kernel factor $\langle q|\hat{\mathrm{T}}_{M,n}|q'\rangle$ leads to the original phase space function and is validated. 

\subsection{Derivation of the leading quantum correction factor for the case of a quartic anharmonic oscillator potential}\label{app:quarticcorrections}

The leading quantum correction for the quartic anharmonic oscillator potential $V(q)=\lambda q^4$ is given by
\begin{equation}\label{tnqpfinalquartn1}
\begin{split}
\hbar^{2}\tau_{M,1}(q,p)&=-\frac{\mu\hbar^{2}}{24}\,\int_{0}^{q}dq'\,\mathrm{exp}\left[\big(V(q)-V(q')\big)\frac{\mu}{p}\frac{\partial}{\partial p}\right]\frac{1}{p} \frac{\partial^{3} \, V(q')}{\partial q'^{3}}\, \frac{\partial^{3} \, \tau_{M,0}(q',p)}{\partial p^{3}}.
\end{split}
\end{equation}

We substitute the LTOA given by Eq. (\ref{ltoaquartic}), expand the exponential function, and perform a term-by-term integration. We get the following expansion

\begin{equation}
 \hbar^{2}\tau_{M,1}(q,p) =\frac{\mu^2 \lambda q^3 \hbar^2}{4} \sum_{k=0}^\infty\, \sum_{l=0}^\infty \frac{(\mu\lambda)^k\,q^{4k}(-2\mu\lambda q^4)^l\,(1/2))_l}{(5/4)_l} \frac{\Gamma(l+3/4)}{\Gamma(k+l+7/4)} \left(\frac{1}{p}\frac{\partial}{\partial p}\right)^k\left(\frac{1}{p}\frac{\partial^3}{\partial p^3}\right)^k p^{-2l-1}.
\end{equation}

The partial derivatives along $p$ is evaluated using the following result

\begin{equation}
\left(\frac{1}{p}\frac{\partial}{\partial p}\right)^k p^{-n}=(-1)^k \frac{(n+2k-2)!!}{(n-2)!!}p^{-n-2k}.
\end{equation}
Thus, we arrive at
\begin{equation}
 \hbar^{2}\tau_{M,1}(q,p) =-\frac{\mu^2 \lambda q^3 \hbar^2}{4\,p^5} \sum_{k=0}^\infty\, \sum_{l=0}^\infty \frac{(-\mu\lambda q^4)^{k+l}\,(2l+2)\,(2l+2k+3)!!}{(5/4)_l} \frac{\Gamma(l+3/4)}{\Gamma(k+l+7/4)} \, p^{-2k-2l}.
\end{equation}
We can decouple the two infinite series by interchanging the order of summations in accordance to Eq. (\ref{intersumm}). Hence, we arrive at Eq. (\ref{2ndcorrecquart}). 

Performing the indicated sum along $k$, we also get
\begin{equation}
\begin{split}
 \hbar^{2}\tau_{M,1}(q,p) &=\frac{\mu^2 \lambda q^3 \hbar^2}{2\,p^5} \sum_{k=0}^\infty  \frac{\Gamma(3/4)\Gamma(k+5/2)}{\sqrt{\pi}\,\Gamma(k+7/4)}  \left(-\frac{2\mu\lambda q^4}{p^2}\right)^k \\
 &- \frac{\mu^2 \lambda q^3 \hbar^2}{6\,p^5} \sum_{k=0}^\infty  \frac{(4k+5)\,(2k+5)\,\Gamma(5/4)\,\Gamma(k+5/2)\,\Gamma(k+7/4)}{\sqrt{\pi}\,\Gamma(k+7/4)\,\Gamma(k+9/4)}  \left(-\frac{2\mu\lambda q^4}{p^2}\right)^k.
 \end{split}
\end{equation}
For $2\mu\lambda q^4/p^2 <1$, the above infinite series converges to $\mathcal{T}_1(q,p,\hbar^2)$ given by Eq. \eqref{t1qpquartic}. 

\begin{thebibliography}{100}

\bibitem{Muga2008} G. Muga, R.S. Mayato, I. Egusquiza, eds., Time in Quantum Mechanics, Vol. 1 (Springer, Berlin, Heidelberg, 2008). \url{https://doi.org/10.1007/978-3-540-73473-4}

\bibitem{Muga2009} G. Muga, A. Ruschhaupt, A. del Camo, eds., Time in Quantum Mechanics, Vol. 2 (Springer-Verlag Berlin Heidelberg, 2009). \url{https://doi.org/10.1007/978-3-642-03174-8}

\bibitem{Galapon2001}E.A. Galapon, Quantum-classical correspondence of dynamical observables, quantization, and the time of arrival correspondence problem. {\em Opt. Spectrosc.} \textbf{91}, 399--405 (2001). \url{https://doi.org/10.1134/1.1405219}

\bibitem{Galapon2002}E.A. Galapon, Pauli's theorem and quantum canonical pairs: the consistency of a bounded, self-–adjoint time operator canonically conjugate to a Hamiltonian with non–empty point spectrum. {\em Proc. R. Soc. Lond. A} \textbf{458}, 451--472 (2002). \url{https://doi.org/10.1098/rspa.2001.0874}

\bibitem{Eckle2008}P. Eckle, A. N. Pfeiffer, C. Cirelli, A. Staudte, R. Dörner, H. G. Muller , M. Büttiker, U. Keller, Attosecond ionization and tunneling delay time measurements in Helium. {\em Science} \textbf{322}, 1525--1529 (2008). \url{https://doi.org/10.1126/science.1163439}

\bibitem{Sainadh2019}U. S. Sainadh, H. Xu, X. Wang, A. Atia-Tul-Noor, W.C. Wallace, N. Douguet, A. Bray, I. Ivanov, K. Bartschat, A. Kheifets, R.T. Sang, I.V. Litvinyuk, Attosecond angular streaking and tunnelling time in atomic hydrogen. {\em Nature} \textbf{568}, 75--77 (2019). \url{https://doi.org/10.1038/s41586-019-1028-3}

\bibitem{Eckle2008a}P. Eckle, M. Smolarski, P. Schlup, J. Biegert, A. Staudte, M. Sch\"{o}ffler, H. G. Muller, R. Dörner, U. Keller, Attosecond angular streaking. {\em Nature Phys} \textbf{4}, 565--570 (2008). \url{https://doi.org/10.1038/nphys982}

\bibitem{Allcock1969a}G.R. Allcock, The time of arrival in quantum mechanics I. Formal considerations. {\em Ann. Phys.} \textbf{53}, 253--285 (1969). \url{https://doi.org/10.1016/0003-4916(69)90251-6}

\bibitem{Allcock1969b}G.R. Allcock, The time of arrival in quantum mechanics II. The individual measurement. {\em Ann. Phys.} \textbf{53}, 286--310 (1969). \url{https://doi.org/10.1016/0003-4916(69)90252-8}

\bibitem{Allcock1969}G.R. Allcock, The time of arrival in quantum mechanics III. The measurement ensemble. {\em Ann. Phys.} \textbf{53}, 311--348 (1969). \url{https://doi.org/10.1016/0003-4916(69)90253-X}

\bibitem{Grot1996}N. Grot, C. Rovelli, R.S. Tate, Time of arrival in quantum mechanics. {\em Phys. Rev. A} \textbf{54}, 4676--4690 (1996). \url{https://doi.org/10.1103/PhysRevA.54.4676}

\bibitem{Aharonov1998}Y. Aharonov, J. Oppenheim, S. Popescu, B. Reznik, W.G. Unruh, Measurement of time of arrival in quantum mechanics. {\em Phys. Rev. A} \textbf{57}, 4130--4139 (1998). \url{https://doi.org/10.1103/PhysRevA.57.4130}

\bibitem{Leavens1998}C.R. Leavens, Time of arrival in quantum and Bohmian mechanics. {\em Phys. Rev. A} \textbf{58}, 840--847 (1998). \url{https://doi.org/10.1103/PhysRevA.58.840}

\bibitem{Delgado1998}V. Delgado, Probability distribution of arrival times in quantum mechanics. {\em Phys. Rev. A} \textbf{57}, 762--770 (1998). \url{https://doi.org/10.1103/PhysRevA.57.762}

\bibitem{Wang2007}Z. Wang, C. Xiong, How to introduce time operator. {\em Ann. Phys.} \textbf{322} 2304--2314 (2007). \url{https://doi.org/10.1016/j.aop.2006.10.007}

\bibitem{Muga2000}J.G. Muga, C.R. Leavens, Arrival time in quantum mechanics. {\em Phys. Rep.} \textbf{338}, 353--438 (2000). \url{https://doi.org/10.1016/S0370-1573(00)00047-8}

\bibitem{J.Leon20000}J. Leon, J. Julve, P. Pitanga, F.J. de Urries, Time of arrival in the presence of interactions. {\em Phys. Rev. A.} \textbf{61}, 062101 (2000). \url{https://doi.org/10.1103/PhysRevA.61.062101}

\bibitem{Galapon2004}E.A. Galapon, Shouldn't there be an antithesis to quantization?. {\em J. Math. Phys.} \textbf{45} 3180--3215 (2004). \url{https://doi.org/10.1063/1.1767297}

\bibitem{Galapon2018}E.A. Galapon, J.J.P. Magadan, Quantizations of the classical time of arrival and their dynamics. {\em Ann. Phys.} \textbf{397}, 278--302 (2018). \url{https://doi.org/10.1016/j.aop.2018.08.005}

\bibitem{Galapon2009a}E.A. Galapon, Theory of quantum arrival and spatial wave function collapse on the appearance of particle. {\em Proc. R. Soc. A.} \textbf{465}, 71--86 (2009). \url{https://doi.org/10.1098/rspa.2008.0278}

\bibitem{Galapon2006}E.A. Galapon, Theory of quantum first time of arrival via spatial confinement I: Confined time of arrival operators for continuous potentials. {\em Int. J. Mod. Phys. A} \textbf{21}, 6351--6381 (2006). \url{https://doi.org/10.1142/S0217751X06034215}

\bibitem{Halliwell2015}J.J. Halliwell, J. Evaeus, J. London, Y. Malik, A self–adjoint arrival time operator inspired by measurement models. {\em Phys. Lett. A} \textbf{379}, 2445--2451 (2015). \url{https://doi.org/10.1016/j.physleta.2015.07.040}

\bibitem{Pollak2017}E. Pollak, Transition path time distribution, tunneling times, friction, and uncertainty. {\em Phys. Rev. Lett.} \textbf{118}, 070401 (2017). \url{https://doi.org/10.1103/PhysRevLett.118.070401}

\bibitem{Sombillo2018}D.L.B. Sombillo, E.A. Galapon, Barrier-traversal-time operator and the time-energy uncertainty relation. {\em Phys. Rev. A.} \textbf{97}, 062127 (2018). \url{https://doi.org/10.1103/PhysRevA.97.062127}

\bibitem{Galapon2004a}E.A. Galapon, R.F. Caballar, R.T. Bahague Jr., Confined quantum time of arrivals. {\em Phys. Rev. Lett.} \textbf{93}, 180406 (2004). \url{https://doi.org/10.1103/PhysRevLett.93.180406}

\bibitem{Anastopoulos2006}C. Anastopoulos, N. Savvidou, Time-of-arrival probabilities and quantum measurements. {\em J. Math. Phys.} \textbf{47}, 122106 (2006). \url{https://doi.org/10.1063/1.2399085}

\bibitem{Sombillo2016}D.L.B. Sombillo, E.A. Galapon, Particle detection and non-detection in a quantum time of arrival measurement. {\em Ann. Phys.} \textbf{364}, 261--273 (2016), \url{https://doi.org/10.1016/j.aop.2015.11.008}

\bibitem{Das2021}S. Das, M. Nöth, Times of arrival and gauge invariance. {\em Proc. R. Soc. A} \textbf{477}, 20210101 (2021). \url{https://doi.org/10.1098/rspa.2021.0101}

\bibitem{Galapon2002a}E.A. Galapon, Self–adjoint time operator is the rule for discrete semi–bounded Hamiltonians. {\em Proc. R. Soc. Lond. A} \textbf{458}, 2671--2689 (2002). \url{https://doi.org/10.1098/rspa.2002.0992}

\bibitem{Galapon2005}E.A. Galapon, F. Delgado, J.G. Muga, I. Egusquiza, Transition from discrete to continuous time-of-arrival distribution for a quantum particle. {\em Phys. Rev. A} \textbf{72}, 042107 (2005). \url{https://doi.org/10.1103/PhysRevA.72.042107}

\bibitem{Galapon2005a}E.A. Galapon, R.F. Caballar, R. Bahague Jr., Confined quantum time of arrival for the vanishing potential. {\em Phys. Rev. A} \textbf{72}, 062107 (2005). \url{https://doi.org/10.1103/PhysRevA.72.062107}

\bibitem{Galapon2008}E.A. Galapon, A. Villanueva, Quantum first time-of-arrival operators. {\em J. Phys. A Math. Theor.} \textbf{41}, 455302 (2008). \url{https://doi.org/10.1088/1751-8113/41/45/455302}

\bibitem{Caballar2009}R.F. Caballar, E.A. Galapon, Characterizing multiple solutions to the time-energy canonical commutation relation via quantum dynamics. {\em Phys. Lett. A} \textbf{373}, 2660--2666 (2009). \url{https://doi.org/10.1016/j.physleta.2009.05.068}

\bibitem{Caballar2010}R.F. Caballar, L.R. Ocampo, E.A. Galapon, Characterizing multiple solutions to the time-energy canonical commutation relation via internal symmetries. {\em Phys. Rev. A} \textbf{81}, 062105 (2010). \url{https://doi.org/10.1103/PhysRevA.81.062105}

\bibitem{Villanueva2010}A.D. Villanueva, E.A. Galapon, Generalized crossing states in the interacting case: The uniform gravitational field. {\em Phys. Rev. A} \textbf{82}, 052117 (2010). \url{https://doi.org/10.1103/PhysRevA.82.052117}

\bibitem{Flores2019}P.C.M. Flores, E.A. Galapon, Quantum free-fall motion and quantum violation of the weak equivalence principle. {\em Phys. Rev. A} \textbf{99}, 042113 (2019). \url{https://doi.org/10.1103/PhysRevA.99.042113}

\bibitem{Galapon2009}E.A. Galapon, Quantum wave-packet size effects on neutron time-of-flight spectroscopy. {\em Phys. Rev. A} \textbf{80}, 030102 (2009). \url{https://doi.org/10.1103/PhysRevA.80.030102}

\bibitem{Galapon2012}E.A. Galapon, Only above barrier energy components contribute to barrier traversal time. {\em Phys. Rev. Lett.} \textbf{108}, 170402 (2012). \url{https://doi.org/10.1103/PhysRevLett.108.170402}

\bibitem{Pablico2020}D.A.L. Pablico, E.A. Galapon, Quantum traversal time across a potential well. {\em Phys. Rev. A.} \textbf{101}, 022103 (2020). \url{https://doi.org/10.1103/PhysRevA.101.022103}

\bibitem{Sombillo2014}D.L.B. Sombillo, E.A. Galapon, Quantum traversal time through a double barrier. {\em Phys. Rev. A.} \textbf{90}, 032115 (2014). \url{https://doi.org/10.1103/PhysRevA.90.032115}

\bibitem{Baute2000}A.D. Baute, R.S. Mayato, J.P. Palao, J.G. Muga, L. Egusquiza, Time-of-arrival distribution for arbitrary potentials and Wigner's time-energy uncertainty relation. {\em Phys. Rev. A} \textbf{61}, 022118 (2000). \url{https://doi.org/10.1103/PhysRevA.61.022118}

\bibitem{Madrid2005}R. de la Madrid, The role of the rigged Hilbert space in quantum mechanics. {\em Eur. J. Phys.} \textbf{26}, 287 (2005). \url{https://doi.org/10.1088/0143-0807/26/2/008}

\bibitem{Madrid2002}R. de la Madrid, Rigged Hilbert space approach to the Schrödinger equation. {\em J. Phys. A: Math. Gen.} \textbf{35}, 319--342 (2002). \url{https://doi.org/10.1088/0305-4470/35/2/311}

\bibitem{Pablico2023}D.A.L. Pablico and E.A. Galapon. {\em EPJ Plus}, \textbf{138}, 153 (2023). \url{https://doi.org/10.1140/epjp/s13360-023-03774-z}

\bibitem{Sombillo2012}D.L.B. Sombillo, E.A. Galapon, Quantum time of arrival Goursat problem. {\em J. Math. Phys.} \textbf{53} 043702 (2012). \url{https://doi.org/10.1063/1.3699175}

\bibitem{Groenewold1946}H.J. Groenewold, On the principles of elementary quantum mechanics. {\em Physica} \textbf{12}, 405--460 (1946). \url{https://doi.org/10.1016/S0031-8914(46)80059-4}

\bibitem{Hove1951}L. van Hove, Sur le problème des relations entre les transformations unitaires de la mécanique quantique et les transformations canoniques de la mécanique classique. {\em Belg. Bull. Cl. Sci.} \textbf{37}, 610--620 (1951). \url{https://www.persee.fr/doc/barb_0001-4141_1951_num_37_1_70660}

\bibitem{Gotay1996}M.J. Gotay, H.B. Grundling, G.M. Tuynman, Obstruction results in quantization theory. {\em J. Nonlinear Sci.} \textbf{6} 469--498 (1996). \url{https://doi.org/10.1007/BF02440163}

\bibitem{Gotay1999}M.J. Gotay, J. Grabowski, H.B. Grundling, An obstruction to quantizing compact symplectic manifolds. {\em Proc. Am. Math. Soc.} \textbf{128}, 237--243 (1999). \url{https://www.jstor.org/stable/119406}

\bibitem{Zachos2001}C.K. Zachos, Deformation quantization: Quantum mechanics lives and works in phase-space. {\em Int. J. Mod. Phys. A} \textbf{17}, 297--316 (2001).\url{https://doi.org/10.1142/S0217751X02006079}

\bibitem{Zachos2005}C.K. Zachos, D.B. Fairlie, T.L. Curtright, eds., Quantum Mechanics in Phase Space, Vol. 34 (World Scientific Publishing Co. Pte. Ltd., 2005). \url{https://doi.org/10.1142/5287}

\bibitem{HaiWoongLee1995} H. Lee, Theory and application of the quantum phase-space distribution functions. {\em Phys. Rep.} \textbf{259}, 147--211 (1995). \url{https://doi.org/10.1016/0370-1573(95)00007-4}.

\bibitem{Kim1991}Y. S. Kim, M. E. Noz, Phase space picture of quantum mechanis. (World Scientific Publishing, 1991). \url{https://doi.org/10.1142/1197}.

\bibitem{Farrales2022}R.A.E. Farrales, H.B. Domingo, E.A. Galapon, Conjugates to one particle Hamiltonians in 1-dimension in differential form. {\em Eur. Phys. J. Plus} \textbf{137}, 830 (2022). \url{https://doi.org/10.1140/epjp/s13360-022-02956-5}

\bibitem{Cohen2017}L. Cohen, The eigenvalue problem in phase space. {\em J. Comput. Chem.} \textbf{39}, 1059--1067 (2017). \url{https://doi.org/10.1002/jcc.24884}

\bibitem{Moyal1949a}J.E. Moyal, Quantum mechanics as a statistical theory. {\em Math. Proc. Camb. Philos. Soc.} \textbf{45}, 99--124 (1949). \url{https://doi.org/10.1017/S0305004100000487}

\bibitem{Bayen1978}F. Bayen, M. Flato, C. Fronsdal, A. Lichnerowicz, D. Sternheimer, Deformation theory and quantization. I. Deformations of symplectic structures. {\em Ann. Phys.} \textbf{111}, 61--110 (1978). \url{https://doi.org/10.1016/0003-4916(78)90224-5}

\bibitem{Bayen1977}F. Bayen, M. Flato, C. Fronsdal, A. Lichnerowicz, D. Sternheimer, Quantum mechanics as a deformation of classical mechanics. {\em Lett. Math. Phys.} \textbf{1}, 521–-530 (1977). \url{https://doi.org/10.1007/BF00399745}


\bibitem{Tosiek2012}J. Tosiek, P. Brzykcy, States in the {H}ilbert space formulation and in the phase space formulation of quantum mechanics. {\em Ann. Phys.} \textbf{332}, 1--15 (2012). \url{https://doi.org/10.1016/j.aop.2013.01.010}

\bibitem{MaciejBlaszak2012}M. Błaszak, Z. Domański, Phase space quantum mechanics. {\em Ann. Phys.} \textbf{327}, 167--211, (2012). \url{https://doi.org/10.1016/j.aop.2011.09.006}

\bibitem{Jordan1961}T. F. Jordan, E.C.G. Sudarshan, Lie Group Dynamical Formalism and the Relation between Quantum Mechanics and Classical Mechanics. {\em Rev. Mod. Phys.} \textbf{33}, 515, (1961). \url{https://doi.org/10.1103/RevModPhys.33.515}

\bibitem{Gelfand1964} I.M. Gel'fand, G.E. Shilov, Generalized Functions, Vol. 1. (Academic Press, London, 1964)

\bibitem{case2008}W.B. Case, Wigner functions and Weyl transforms for pedestrians. {\em Am. J. Phys.} \textbf{10}, 937–-946, (2008). \url{https://doi.org/10.1119/1.2957889}

\bibitem{derivativepfq}{Symbolic differentiation of the HypergeometricPFQ[$\{a_1,...,a_p\},\{b_1,...,b_q\},z$] 
 function with respect to $z$, \url{http://functions.wolfram.com/07.31.20.0013.01
} [Accessed 03-June-2024]}.

\bibitem{intpfq}{Integration of the HypergeometricPFQ[$\{a_1,...,a_p\},\{b_1,...,b_q\},z$] 
 function involving power function with respect to $z$, \url{http://functions.wolfram.com/07.31.21.0002.01
} [Accessed 03-June-2024]}.

\bibitem{galapsbarrier}E.A. Galapon, Only above barrier energy components contribute to barrier traversal time. {\em Phys. Rev. A} \textbf{108}, 170402. (2012). \url{https://doi.org/10.1103/PhysRevLett.108.170402}

\bibitem{arxivflores} P.C.M. Flores, D.A.L. Pablico, E.A. Galapon, Instantaneous tunneling time within the theory of time-of-arrival operators. arXiv:2409.12389 [quant-ph], (2024). \url{https://arxiv.org/abs/2409.12389}

\bibitem{ecklehelium}P. Eckle, A.N. Pfeiffer, C. Cirelli, A. Staudte, R. D\"{o}rner, H.G. Muller, M. B\"{u}ttiker, U. Keller, Attosecond Ionization and Tunneling Delay Time Measurements in Helium. {\em Science} \textbf{322}, 1525--1529 (2008). \url{https://doi.org/10.1126/science.1163439}

\bibitem{sainadh}U.S. Sainadh, H. Xu, X. Wang,  A. Atia-Tul-Noor, W.C. Wallace, N. Douguet, A. Bray, I. Ivanov, K. Bartschat, A. Kheifets, R. T. Sang, I. V. Litvinyuk,  Attosecond angular streaking and tunnelling time in atomic hydrogen. {\em Nature} \textbf{568}, 75-–77 (2019). \url{https://doi.org/10.1038/s41586-019-1028-3}


\end{thebibliography}

\end{document}